\documentclass[11pt]{article}

\usepackage[english]{babel}
\usepackage[utf8]{inputenc}
\usepackage{johd}
\usepackage{fancyhdr}
\usepackage{amsmath}
\usepackage{mathrsfs}
\usepackage{xcolor}

\newcommand{\dbar}[1]{\bar{\bar{#1}}}

\title{\textbf{Multimode non-Hermitian framework for third harmonic generation in nonlinear photonic systems comprising 2D materials}}

\author{Thomas Christopoulos$^{a,b,*}$, Emmanouil E. Kriezis$^{b}$, and Odysseas Tsilipakos$^{c,a,*}$ \\
        \small $^{a}$Institute of Electronic Structure and Laser, Foundation for Research and Technology Hellas (FORTH), \\
        \small       Heraklion 70013, Crete, Greece \\
        \small $^{b}$School of Electrical and Computer Engineering, Aristotle University of Thessaloniki (AUTH), \\ 
        \small       Thessaloniki 54124, Greece \\
        \small $^{c}$Theoretical and Physical Chemistry Institute, National Hellenic Research Foundation, \\
        \small       Athens 11635, Greece. \\
        \small $^{*}$Corresponding authors: \tt{cthomasa@iesl.forth.gr}; \tt{otsilipakos@eie.gr}
}

\date{} 

\begin{document}

\maketitle
\thispagestyle{empty}

\begin{abstract} 
    Resonant structures in modern nanophotonics are non-Hermitian (leaky and lossy), and support quasinormal modes. Moreover, contemporary cavities frequently include 2D materials to exploit and resonantly enhance their nonlinear properties or provide tunability. Such materials add further modeling complexity due to their infinitesimally thin nature and strong dispersion. Here, a formalism for efficiently analyzing third harmonic generation (THG) in nanoparticles and metasurfaces incorporating 2D materials is proposed. It is based on numerically calculating the quasinormal modes in the nanostructure, it is general, and does not make any prior assumptions regarding the number of resonances involved in the conversion process, in contrast to conventional coupled-mode theory approaches in the literature. The capabilities of the framework are showcased via two selected examples: a single scatterer and a periodic metasurface incorporating graphene for its high third-order nonlinearity. In both cases, excellent agreement with full-wave nonlinear simulations is obtained. The proposed framework may constitute an invaluable tool for gaining physical insight into the frequency generation process in nano-optic structures and providing guidelines for achieving drastically enhanced THG efficiency.
\end{abstract}


\section{\label{sec:Intro} Introduction}

    Nonlinear light-matter interactions are an indispensable ingredient in contemporary optical systems since they allow for advanced functionalities \cite{Taghinejad:2019,Li:2017}, well beyond the reach of linear phenomena. Harmonic generation, parametric amplification, frequency mixing, multiphoton and saturable absorption, self- and cross-phase modulation are only some of the most notable nonlinear effects \cite{BoydBook}. Nonlinear optics first flourished in bulk crystals and optical fibers where long interaction lengths are available. The recent shift towards compact nanophotonic systems and ultrathin metasurfaces limits the interaction length/volume between the optical field and the nonlinear material necessitating different physical approaches in order to achieve strong nonlinear effects. More specifically, these include (i)~the exploitation of resonant structures with high quality factors to confine energy temporally \cite{Christopoulos:2019OpEx} and small mode volume to confine energy spatially producing enhanced local fields \cite{Sauvan:2013}, and (ii)~the utilization of highly nonlinear materials, be it bulk (nonlinear polymers, chalcogenide glasses, etc.) or, more recently, the emerging category of sheet, two-dimensional (2D) materials \cite{Autere:2018,Guo:2019}. Sheet materials [graphene, transition metal dichalcogenides (TMDs), black phosphorus, MXenes, etc.] have illustrated great potential for photonic and optoelectronic applications \cite{Mak2016,Sinatkas2021,Matthaiakakis:2022}. However, they also introduce complexity when incorporated in a nanophotonic system, both in terms of its practical realization, as well as its efficient analysis and design, e.g., their atomic thickness and strong dispersion needs to be handled carefully.

    Here, we focus on how the two-dimensional nature as well as the lossy and dispersive properties of 2D materials should be handled rigorously in the context of modal analysis tools for open (leaky) resonant third-order nonlinear systems. Note that practically all modern nanophotonic systems, including dielectric/plasmonic particles, periodic metasurfaces, photonic crystal membranes, and compact guided-wave resonators, exhibit significant radiation leakage. Such contemporary systems can be efficiently studied using modal techniques; a research direction which has gathered significant interest recently \cite{Lalanne:2018,Kristensen:2020,Wu:2021,Sauvan:2022,Both:2022}. Initially, researchers focused on developing a \emph{linear} framework capable of handling non-Hermitian resonant structures (leaky and lossy) comprised of \emph{bulk} materials. The supported quasinormal eigenmodes (QNMs) diverge in space away from the resonator and their correct normalization became the subject of numerous studies \cite{Sauvan:2013,Doost:2014,Muljarov:2016,Weiss:2017,Gras:2019}. Building upon this normalization, a number of techniques to reconstruct the full spectrum of a system using the supported QNMs were presented \cite{Muljarov:2016GoldSand,Yan:2018,Kongsuwan:2020,Primo:2020,Zhou:2022,Ren:2022,Ren:2022b}. 
    Other important aspects have been discussed as well, such as completeness and orthogonality of the expansion \cite{Kristensen:2020,Sauvan:2022,Both:2022}, and the importance of including the static modes \cite{Lobanov:2019,Sauvan:2021}. Furthermore, related classical theoretical tools of Hermitian analysis, such as perturbation theory  \cite{Yang:2015NanoLett,Both:2019,Yan:2020,Christopoulos:2020} 
    and temporal coupled-mode theory (CMT) \cite{Christopoulos:2020,Zhang:2020,Benzaouia:2021,Zhou:2021} were modified to become applicable to QNMs and non-Hermitian systems.
    
    Despite the progress concerning linear systems, very limited focus has been directed to the wider class of nonlinear non-Hermitian systems \cite{Christopoulos:2020,Gigli:2020}. In this work, we contribute towards this important direction by developing a QNM-based, multimode third-order nonlinear framework which allows to include 2D photonic materials with loss and dispersion within the resonant structure. It builds on Ref.~\cite{Gigli:2020}, which discusses bulk materials and second-order nonlinearity. 
    Our contribution is two-fold: (i)~it introduces 2D materials in a linear QNM framework, naturally incorporating their unique infinitesimally-thin nature and dispersive properties, and (ii)~it contributes into the exploitation of their  third-order nonlinear properties for efficient frequency generation, under the same QNM perspective. The developed framework can aid in efficiently analyzing and designing general resonant systems comprising bulk \emph{and} sheet materials, as well as gaining valuable physical insight into the resonances mediating the conversion process and developing design directives for obtaining efficient performance.

\section{\label{sec:Framework} Third-harmonic generation framework for photonic scatterers including 2D materials}

    The proposed nonlinear framework is based on the ability to calculate the response of a resonant system upon a prescribed excitation by using a finite set of QNMs \cite{Lalanne:2018, Both:2022}. The first step towards this goal is to correctly calculate and normalize the supported QNMs, a non-trivial task in non-Hermitian systems \cite{Sauvan:2013,Doost:2014,Muljarov:2016,Weiss:2017,Gras:2019} due to the fact that the mode profile outside a resonant cavity diverges \cite{Christopoulos:2019OpEx}. To computationally calculate and normalize the supported QNMs, we build upon Refs.~\cite{Yan:2018,Raman:2011},  extending the methodology in order to tackle contemporary photonic structures including 2D and bulk materials with lossy and dispersive electomagnetic properties.
    
    For brevity, below we present the derivation concerning the 2D-material inclusions; the complete case can be found in App.~\ref{sec:GeneralFramework}. Henceforth, a 2D material with a Drude-like complex surface conductivity (measured in $\mathrm{S}$) of the form $\dbar{\sigma}_s(\omega) = -j\dbar{\sigma}_0/(\omega-j\gamma)$ is assumed. It is convenient to express the source-free curl Maxwell's equations using the compact notation $\hat{\mathcal{L}}\mathbf{\tilde\Psi}_m = \tilde\omega_m\mathbf{\tilde\Psi}_m$, where $\mathbf{\tilde\Psi}_m = [\mathbf{\tilde H}_m~~\mathbf{\tilde E}_m~~\mathbf{\tilde J}_m]^T$ is a supervector containing the involved electromagnetic fields, $m$ is a general index of the QNM order, $\hat{\mathcal{L}}$ is a curl operator (in matrix notation)
    \begin{equation}
        \hat{\mathcal{L}} = \begin{bmatrix}  0                              & j\mu^{-1}\nabla\times    & 0                \\ 
                                             -j\varepsilon^{-1}\nabla\times & 0                        & -\varepsilon^{-1} \\
                                             0                              & -\dbar{\sigma}_0\delta_s & j\gamma
                            \end{bmatrix},
        \label{Eq:LLinear}
    \end{equation}
    $\mathbf{\tilde J}_m = -\dbar{\sigma}_0/(\tilde\omega_m-j\gamma) \mathbf{\tilde E}_m\delta_s$ is a surface auxiliary field that is used to introduce the dispersive nature of the 2D material, and $\delta_s\equiv\delta_s(\mathbf{r})$ is a surface Dirac function used to capture the presence of the 2D material. Note that $\mathbf{\tilde J}_m$ does not correspond to the (surface) current density $\mathbf{J}_s = \dbar{\sigma}_s\mathbf{E}$ of the Maxwell's equations (they differ by a prefactor $j$).
    The tilde in the fields denotes modal quantities. Similarly, $\tilde\omega_m$ is the respective complex eigenvalue of the $m$-th order QNM; the imaginary part  carries information regarding the linewidth of the mode. The second rank tensor $\dbar{\sigma}_0$ is introduced to encapsulate the 2D nature of the involved material in the sense that the field interacts with the sheet material only through its tangential field components. Finally, $\gamma$ describes damping due to Ohmic loss. 
    
    It has been shown that due to the orthonormality of QNMs \cite{Yan:2018} the scattered field of a resonator can be expanded into an infinite sum of the form $\mathbf{\Psi}_\mathrm{sct}(\omega) = \sum_m a_m(\omega) \mathbf{\tilde \Psi}_m$, where the expansion coefficients $a_m(\omega)$ are calculated through (see App.~\ref{sec:GeneralFramework})
    \begin{equation}
        a_m(\omega) = \frac{1}{\tilde\omega_m - \omega}\int_V \mathbf{\tilde J}_m \cdot \mathbf{E}_\mathrm{inc} \mathrm{d}V, \label{Eq:alphaLinear}
    \end{equation}
    where $\mathbf{E}_\mathrm{inc}$ is the incident field that excites the resonant scatterer at the real frequency $\omega$ and must include possible reflections from the background \cite{Gras:2019}. Equation~\eqref{Eq:alphaLinear} applies only when a (surface) conductivity term is present; for the more general case of bulk \emph{and} sheet materials see App.~\ref{sec:GeneralFramework}. Although the completeness of the expansion is not strictly ensured for any arbitrary geometry, it is widely accepted that the expansion is complete inside and in the vicinity of the resonant cavity \cite{Sauvan:2022,Both:2022}. In Eq.~\eqref{Eq:alphaLinear}, all the involved modes are normalized such that
    \begin{equation}
        \int_V \left( \mathbf{\tilde E}_m\cdot\varepsilon\mathbf{\tilde E}_m - \mathbf{\tilde H}_m\cdot\mu\mathbf{\tilde H}_m - \mathbf{\tilde E}_m\cdot j\left.\frac{\partial\dbar{\sigma}_s(\omega)}{\partial\omega}\right|_{\omega = \tilde\omega_m}\mathbf{\tilde E}_m\delta_s \right) \mathrm{d}V = 1. \label{Eq:QNMsNormalization}
    \end{equation}
    Equations~\eqref{Eq:alphaLinear}~and~\eqref{Eq:QNMsNormalization} are key ingredients in the description of linear resonant systems incorporating dispersive 2D materials with the QNM framework. The last term on the left-hand side of Eq.~\eqref{Eq:QNMsNormalization} is an important and necessary addition, providing the correct normalization of the QNMs when the nanophotonic resonator includes 2D materials; recently, we have revealed a similar contribution in the normalization for Hermitian (or quasi-Hermitian) systems \cite{Christopoulos:2016PRE}. Omitting this term introduces large errors \cite{Christopoulos:2016PRE}, especially when dispersion [see spectral derivative in Eq.~\eqref{Eq:QNMsNormalization}] is strong (see also App.~\ref{subsec:QNMsNormalization} for a comparison). Note that in order to correctly evaluate Eq.~\eqref{Eq:QNMsNormalization} one has to terminate the computational domain with perfectly-matched layers (PMLs) and perform the integration inside the artificial domains as well \cite{Sauvan:2013}. Alternatively, an appropriate surface term should be added to truncate the infinite space into a finite computational domain \cite{Doost:2014}. In this case, the applicability is limited to systems with a uniform background material \cite{Sauvan:2022}.

    A similar approach can be followed to describe \emph{nonlinear} effects, as shown in Ref.~\cite{Gigli:2020} for the case of bulk materials and second-order nonlinearity. In this work, we allow for the inclusion of 2D materials and focus on third-order nonlinearities and third harmonic generation (THG) in particular. Note that coupled-mode theory frameworks involving a single mode in each of the fundamental and the third-harmonic frequencies exist in the literature \cite{Rodriguez:2007,Hashemi:2009}. However, they fail to capture the correct physical picture when more resonances are involved in the conversion process. In sharp contrast, the proposed  \emph{multimode} framework allows for interaction between all the supported QNMs without making any prior assumptions regarding which modes mediate the conversion process. We start by writing the scattered fields at the fundamental and third harmonic frequencies as
    \begin{subequations}
        \begin{equation}
            \hat{\mathcal{L}}\mathbf{\Psi}_\mathrm{sct}^{(\omega)} = \omega\mathbf{\Psi}_\mathrm{sct}^{(\omega)} + \mathbf{S}_\mathrm{inc}^{(\omega)}, \label{Eq:NLScatterOmega}
        \end{equation}    
        \begin{equation}
            \hat{\mathcal{L}}\mathbf{\Psi}_\mathrm{sct}^{(3\omega)} = 3\omega\mathbf{\Psi}_\mathrm{sct}^{(3\omega)} + \mathbf{S}_\mathrm{inc}^{(3\omega)}, \label{Eq:NLScatter3Omega}
        \end{equation}
        \label{Eq:NLScatter}
    \end{subequations}
    where $\mathbf{S}_\mathrm{inc}^{(\omega)} = [0~~0~~\dbar{\sigma}_0\mathbf{E}_\mathrm{inc}\delta_s]^T$ is the direct excitation at $\omega$ (e.g., an appropriately-polarized plane wave) and $\mathbf{S}_\mathrm{inc}^{(3\omega)} = [0~~-j\varepsilon^{-1}\mathbf{J}_{s,\mathrm{NL}}^{(3\omega)}\delta_s~~0]^T$ is the induced nonlinear source at $3\omega$ (no direct external excitation at $3\omega$ takes place). The induced nonlinear surface current $\mathbf{J}_{s,\mathrm{NL}}^{(3\omega)} = (\sigma_3/4)  (\mathbf{E}_{t,\parallel}^{(\omega)} \cdot \mathbf{E}_{t,\parallel}^{(\omega)})\mathbf{E}_{t,\parallel}^{(\omega)}$, acting as the source at the third harmonic frequency, is determined from the nonlinear interaction between the total tangential electric field components at the fundamental frequency, i.e., the sum the sum of the incident, reflected from the background, and scattered fields  ($\mathbf{E}_t^{(\omega)} = \mathbf{E}_\mathrm{inc}^{(\omega)} + \mathbf{E}_r^{(\omega)}  + \mathbf{E}_\mathrm{sct}^{(\omega)} = \mathbf{E}_b^{(\omega)} + \mathbf{E}_\mathrm{sct}^{(\omega)}$) \cite{Cheng:2014}. By substituting the QNM expansion at the fundamental frequency in Eq.~\eqref{Eq:NLScatterOmega} and following a procedure similar to Ref.~\cite{Yan:2018}, we can retrieve Eq.~\eqref{Eq:alphaLinear} (see also App.~\ref{sec:GeneralFramework}). In a completely analogous manner, we substitute the respective expansion at the third harmonic frequency, $\mathbf{\Psi}_\mathrm{sct}^{(3\omega)} = \sum_m a_m(3\omega) \mathbf{\tilde \Psi}_m$, in Eq.~\eqref{Eq:NLScatter3Omega} and arrive at
    \begin{equation}
        a_m(3\omega) = \frac{-j}{\tilde\omega_m - 3\omega}\int_V \mathbf{\tilde E}_m \cdot \mathbf{J}_{s,\mathrm{NL}}^{(3\omega)}\delta_s \mathrm{d}V. \label{Eq:alphaNonlinear}
    \end{equation}
    Equation~\eqref{Eq:alphaNonlinear} constitutes the second and most important result of this work. It allows to calculate the expansion coefficients and, thus, the nonlinear response of a resonant system with third-order nonlinear sheet materials. To do so, one first needs to specify the nonlinear current $\mathbf{J}_{s,\mathrm{NL}}^{(3\omega)}$ using the scattered $\mathbf{E}_\mathrm{sct}^{(\omega)}$ and the background $\mathbf{E}_b^{(\omega)} = \mathbf{E}_\mathrm{inc}^{(\omega)} + \mathbf{E}_r^{(\omega)}$ fields at the fundamental frequency. $\mathbf{E}_\mathrm{sct}^{(\omega)}$ is reconstructed from the eigenmodes and their respective amplitudes $a_m(\omega)$, while $\mathbf{E}_b^{(\omega)}$ is calculated either analytically or numerically in the absence of the resonant cavity. 
    We shall highlight that when $\mathbf{E}_b^{(\omega)} \neq \mathbf{E}_\mathrm{inc}^{(\omega)}$ (i.e., in the presence of a substrate), the former should replace the latter even when $a_m(\omega)$ are calculated.
    Then, the scattered field at the third harmonic $\mathbf{E}_\mathrm{sct}^{(3\omega)}$ can be reconstructed using the same eigenmodes but weighted with the new amplitudes $a_m(3\omega)$ [Eq.~\eqref{Eq:alphaNonlinear}] (now those in the neighborhood of $3\omega$ are expected to mainly contribute). Note that this strategy cannot take into account neither the power lost from the field at the fundamental frequency nor the back-conversion from the third-harmonic to the fundamental frequency. It is termed the \emph{undepleted pump} approximation and is accurate for moderate conversion efficiencies \cite{Christopoulos:2018}.
    
\section{\label{sec:Examples} Evaluation of nonlinear framework in graphene-comprising resonant systems}

\begin{figure}[t]
        \centering
        \includegraphics{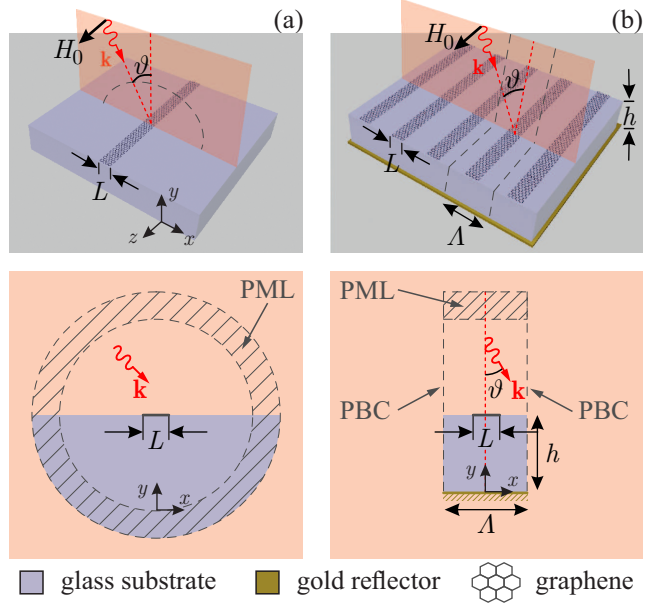}
        \caption{Schematic illustration of the two graphene-based resonant systems under study. (a) ~Single scatterer: a graphene strip of length $L$ resides on a glass substrate. (b)~Periodic system: a metasurface comprised of graphene strips with pitch $\Lambda$ on top of a metal-backed substrate to operate in reflection. The metasurface is designed to act as a perfect absorber for a specific quasinormal mode. TM-polarized incidence inside the $xy$-plane is considered ($\mathbf{H}\equiv H_z\hat{\mathbf{z}}$) in order to excite graphene surface plasmons propagating along the $x$ axis. The bottom panels depict the considered two-dimensional computational domains along with the perfectly matched layers (PMLs) used for window truncation. In order to correctly evaluate Eq.~\eqref{Eq:QNMsNormalization}, one has to perform the integration inside the PMLs as well (see main text).}
        \label{Fig:SystemsSchematic}
    \end{figure}

    To highlight the capabilities of the developed nonlinear framework, we analyze two resonant structures comprising graphene, the most prominent representative of the 2D materials family. We choose THz frequencies  where graphene supports tightly confined surface plasmons (GSPs) \cite{Bludov:2013}.
    The two structures under study are depicted in Fig.~\ref{Fig:SystemsSchematic}. The first is a single scatterer: a graphene strip lying on a glass substrate [Fig.~\ref{Fig:SystemsSchematic}(a)]. The second is a periodic system derived from the first one, i.e., a metasurface made of periodically arranged graphene strips on a metal-backed substrate to operate in reflection [Fig.~\ref{Fig:SystemsSchematic}(b)]. Transverse-magnetic (TM) polarized incidence in the $xy$-plane is considered in order to excite graphene surface plasmons propagating along the $x$ axis. In what follows, the response of both systems around the third harmonic frequency is studied using the developed QNM framework and validated through nonlinear full-wave simulations \cite{Christopoulos:2018}.

\subsection{\label{subsec:ExamplesSingle} Graphene-strip single scatterer on a glass substrate}

\begin{figure*}[t]
        \centering
        \includegraphics{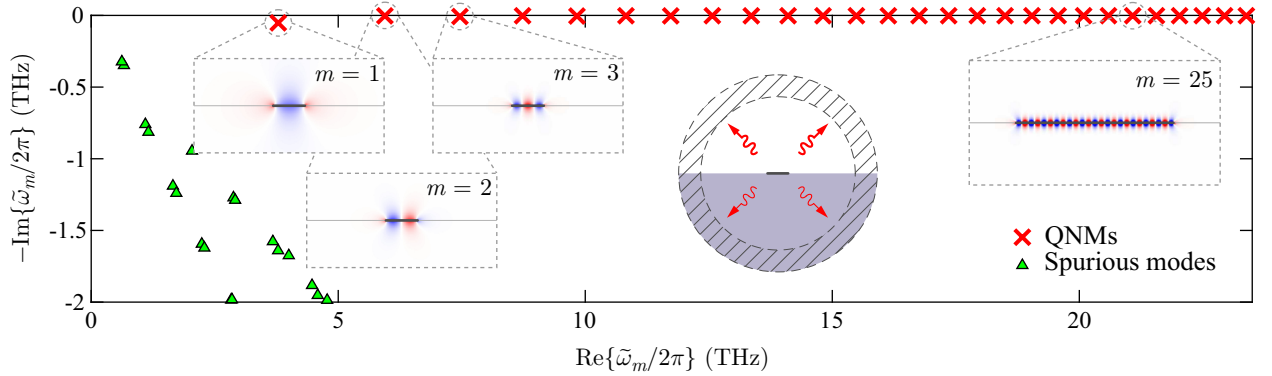}
        \caption{Spectrum of QNMs and spurious modes for the examined graphene-strip scatterer. Red ``X'' markers designate the QNMs. They all have small imaginary parts indicating high quality factors. They are unevenly spaced along the real axis due to graphene dispersion. The mode profiles ($E_x$ component) are included as insets for the three lowest-order modes and one higher-order ($m=25$) mode, lying closest to the third harmonic frequency of the $m=3$ mode, which will be exploited for the excitation at the fundamental frequency. Green triangle markers highlight a few spurious modes, which are also included in the expansions to obtain better accuracy.}
        \label{Fig:SingleGrQNMs}
    \end{figure*}

    We first consider a single graphene strip of length $L = 5~\mathrm{\mu m}$, lying on a glass substrate with $n = 1.45$ [Fig.~\ref{Fig:SystemsSchematic}(a)]. Graphene conductivity in the THz frequency band has a strongly dispersive Drude-like behavior; as is the case with metals below the plasma frequency, this allows for supporting strongly confined plasmons \cite{Bludov:2013}. Here, we adopt the  parameters $\gamma = 1/40~\mathrm{Trad/s}$, $\sigma_0 = e^2\mu_c/\pi\hbar^2$, and $\mu_c = 0.3~\mathrm{eV}$ for its linear properties \cite{Hanson:2008JAP} and $\sigma_3 = +j1.2\times10^{-18}~\mathrm{S(m/V)^2}$ to describe the third-order nonlinear response \cite{Cheng:2014}. Using the commercial Finite Element Method software COMSOL Multiphysics\textsuperscript{\textregistered}, we are able to calculate the QNMs of this leaky and dispersive system, building on the auxiliary-fields approach presented in Refs.~\cite{Yan:2018,Raman:2011} and appropriately extending it to  include sheet materials with Drude dispersion such as graphene. A relatively small set of the eigenmode solutions returned by the solver of COMSOL is depicted in Fig.~\ref{Fig:SingleGrQNMs}. All QNMs are clearly marked with a red ``X'' indicator and have a relatively small imaginary part which results in quality factors ranging from a few tens/hundreds (low order modes) to a few thousands (higher order modes). They correspond to standing waves of GSPs propagating along the $x$-axis and getting reflected at the edges of the finite strip, satisfying a Fabry-P\'{e}rot-type resonant condition of the form $2k_{\parallel}^\mathrm{GSP}L+2\phi_r=n 2\pi$, where $\phi_r \approx -3\pi/4$ is the reflection phase at the strip termination \cite{Nikitin:2014}, which is almost dispersionless as we have numerically verified and differs from the expected $-\pi$ reflection phase of, e.g., plasmons in metals. Note that the integer $n$ does not coincide with the mode index $m$, which measures the number of half-wavelengths along the graphene strip. The reflection phase $\phi_r$ and the phase accumulated as the GSP propagates ($k_{\parallel}^\mathrm{GSP}L$) are of opposite signs and the resonance condition of, e.g., the $m=3$ QNM is fulfilled at a total accumulated phase of $4\pi$, i.e., $n=m-1$ here. The Fabry-P\'{e}rot nature of the modes is also verified by the equidistant spacing when momentarily considering a non-dispersive (unphysical case) graphene strip (see Fig.~\ref{Fig:SingleGrQNMsNonDisp}). 
    The field profiles of a few QNMs are included as insets, showing the three lowest-order modes and a higher-order one ($m = 25$). The $m = 25$ mode is the closest to the third harmonic frequency of the $m = 3$ order mode ($\omega_\mathrm{res,3}/2\pi = 7.46~\mathrm{THz}$ and $\omega_\mathrm{res,25}/2\pi = 22.49~\mathrm{THz}$), which is chosen to act as the fundamental mode, i.e., to accommodate the excitation at the fundamental frequency. This is justified by the fact that $m=3$ is the first symmetric (bright) mode with a high quality factor (exceeding 100), thus providing strong resonant enhancement to boost the conversion process. Note that antisymmetric modes cannot be excited with a normally-incident plane wave (see App.~\ref{subsec:QNMsSpectumQreconstruction} for a relevant discussion). Despite being of Fabry-P\'{e}rot type, the QNMs are unevenly spaced, a direct consequence of the highly dispersive conductivity of graphene and, to a much lesser extent, of the dispersion of the reflection phase $\phi_{r}$ experienced by the propagating GSP at the edges of the strip \cite{Nikitin:2014} (see App.~\ref{subsec:QNMsSpectumFull} for the pole structure of the system when material dispersion is momentarily ignored). Thus, we do not find a QNM at exactly $3\omega_\mathrm{res,3}$, but rather one lying in its vicinity, and the respective mode order is $m=25$ instead of $m = 9$.
    
    Spurious modes are also returned by the COMSOL eigenvalue solver and are marked with green triangles in Fig.~\ref{Fig:SingleGrQNMs} (see also Fig.~\ref{Fig:SingleGrQNMsFull} for a more expanded spectrum). These modes, mainly located either inside the PML (PML modes) or spanning the entire computational domain, should be included in the calculations to ensure the best possible accuracy \cite{Yan:2018,Gigli:2020}. For a more in-depth discussion of the PML modes and the accuracy of the linear framework, the reader is referred to Ref.~\cite{Lalanne:2018}.
    
\begin{figure}[!t]
        \centering
        \includegraphics{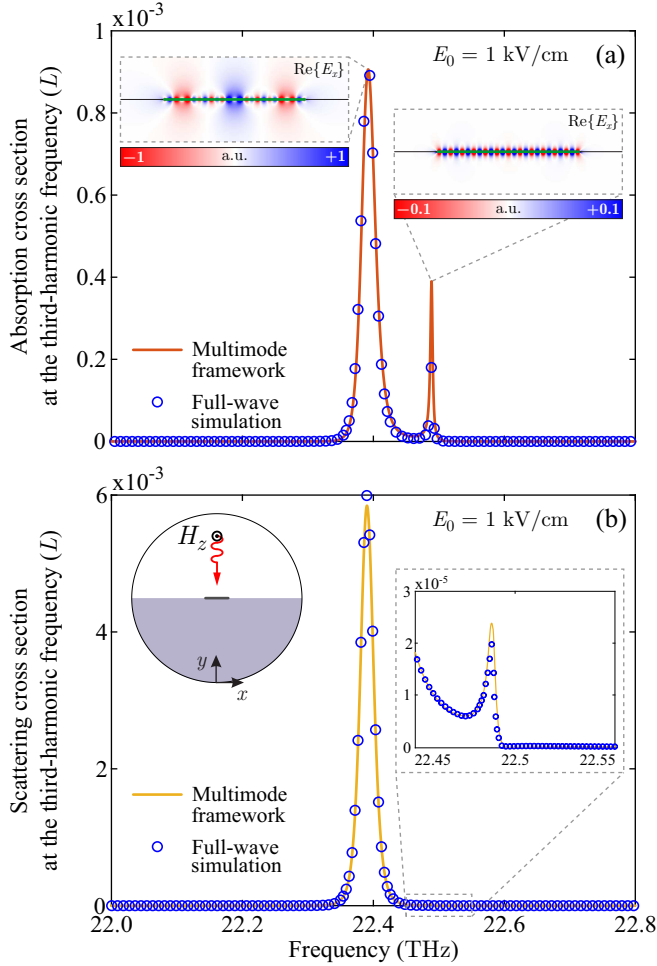}
        \caption{Third harmonic generation with a graphene-strip scatterer on a semi-infinite glass substrate. Evaluation of the proposed nonlinear framework by comparing with full-wave simulations when $E_0 = 1~\mathrm{kV/cm}$. (a)~Absorption cross-section at the third harmonic frequency in units of graphene strip length. Insets: $E_x$-field distribution as obtained from full-wave simulations at the two absorption peaks with the second lying exactly at the resonance frequency of the $m=25$ QNM. (b)~Scattering cross-section at the third harmonic frequency in units of graphene strip length. Inset: Zoomed-in plot around $22.49~\mathrm{THz}$ where the second peak lies. Even weak spectral features are accurately resolved, testifying for the  high accuracy of the proposed framework.}
        \label{Fig:SingleGrNLXsections}
    \end{figure}

    As a first evidence of the capabilities of the proposed framework, in Fig.~\ref{Fig:SingleGrNLXsections} we plot the absorption and scattering cross-sections of the graphene-strip  scatterer (solid lines) at the vicinity of the third harmonic frequency after illumination with a TM-polarized, normally-incident plane wave towards the $-\mathbf{\hat y}$ direction of the form $\mathbf{E}_\mathrm{inc} = E_0\exp{\{+jk_0y\}}\mathbf{\hat x}$ [harmonic time convention: $\exp{\{+j\omega t\}}$] and $E_0 = 1~\mathrm{kV/cm}$ here. The cross-sections are calculated through the equations shown in App.~\ref{subsec:LinearExavulationSinge} by using the scattered field $\mathbf{E}_\mathrm{sct}^{(3\omega)}$ as obtained from the eigenmode expansion utilizing the $a_m(3\omega)$ coefficients [Eq.~\eqref{Eq:alphaNonlinear}] as the weights of the sum. Note that each point on the graph corresponds to a different illumination frequency $\omega$. A direct comparison with the respective results obtained using a rigorous full-wave nonlinear solver (blue circles) reveals the very high accuracy of the developed QNM framework. Due to the continuous wave (CW) nature of the excitation, the nonlinear full-wave simulation can be decomposed in two independent steps (linear simulations). First, a linear scattering problem is solved at $\omega$ to retrieve $\mathbf{E}_\mathrm{sct}^{(\omega)}$. Then, the total field, i.e., the summation of the analytically known background field and the calculated scattered field, is used to calculate the nonlinear surface current, which acts as an induced source for a linear radiation problem at the third harmonic frequency. By solving the second problem, we specify $\mathbf{E}_\mathrm{sct}^{(3\omega)}$ and, thus, we are able to calculate the respective scattering cross-sections included in Fig.~\ref{Fig:SingleGrNLXsections} with markers. 
    
    Interestingly, two distinct peaks appear in the absorption cross-section [Fig.~\ref{Fig:SingleGrNLXsections}(a)]. As mentioned, due to graphene dispersion a higher-order resonance is not found at exactly $3\omega_\mathrm{res,3}$. The simultaneous presence of resonances at exactly $\omega$ and $3\omega$ is frequently termed double-resonant enhancement \cite{Rodriguez:2007,You2017} and would correspond to conditions for optimum conversion efficiency,  Here, this is not exactly the case, leading to two distinct peaks. The first peak emerges precisely at the third harmonic of the fundamental mode, i.e., at $3\omega_\mathrm{res,3}/2\pi = 22.38~\mathrm{THz}$. The second peak lies at the resonance frequency of the $m = 25$ order mode, i.e., at $22.49~\mathrm{THz}$. The fact that the enhanced conversion process at $22.49~\mathrm{THz}$ is mediated by the $m = 25$ order mode is further corroborated by the distribution of the $E_x$ component observed in the full-wave simulations (radiation by nonlinear current on graphene), see inset in Fig.~\ref{Fig:SingleGrNLXsections}(a); it features exactly the same distribution as the pure $m = 25$ QNM extracted from the eigenmode analysis (c.f. Fig.~\ref{Fig:SingleGrQNMs}). The observation of the second peak is a consequence of the low quality factor of the fundamental ($m=3$) mode. Given that $Q_{i,3} = 486.4$, the frequency $22.49/3 = 7.4967~\mathrm{THz}$ still lies under the Lorentzian of the fundamental mode and, thus, a non-negligible amount of light interacts with graphene and is up-converted. Since no QNM is found at exactly $22.38~\mathrm{THz}$, a hybrid field distribution 
    is seen in the corresponding inset.
    The second peak exhibits approximately half the amplitude of the first one due to the sub-optimal coupling of the respective fundamental frequency (note that the field distributions in the insets have an order of magnitude difference). Finally, a secondary peak appears in the scattering cross-section as well [inset of Fig.~\ref{Fig:SingleGrNLXsections}(b)], but with a much lower amplitude. This is attributed to the corresponding resistive and radiative quality factors, which equal $Q_\mathrm{res,25} = 5\;656.7$ and $Q_\mathrm{rad,25} = 89\;766$, respectively, and to the fact that $\sigma_\mathrm{abs} \propto r_Q/(1+r_Q)^2$ while $\sigma_\mathrm{sct} \propto 1/(1+r_Q)^2$, with $r_Q = Q_\mathrm{rad}/Q_\mathrm{res}$ \cite{Christopoulos:2019OpEx,Conteduca:2022}.
    
    We stress that single mode frameworks such as the classical form of the CMT \cite{Christopoulos:2018,Tsilipakos:2016}, are not able to capture such a complex behavior.
    On the contrary, the presented multimode framework captures the actual physical picture with multiple resonances being involved in the conversion process. Finally, note that if we want to shift the resonance positions and place a higher-order resonance exactly at the third harmonic of the fundamental frequency to enhance the conversion efficiency, we can do so by resorting to a finite-width scatterer and tuning the dimension along the $z$-axis. This way, the underlying waveguide becomes of finite width introducing waveguide dispersion and providing an additional degree of freedom in shaping the total dispersion of the propagating GSP and, thus, the positions of the resonances \cite{Theodosi:2021}.
    
\subsection{\label{subsec:ExamplesMeta} Graphene-strip metasurface on a metal-backed glass substrate}

    For the second example, we will switch to a periodic system, i.e., we expand the single graphene strip of the previous section into a metasurface, which is backed by a gold reflector to operate in reflection mode  [Fig.~\ref{Fig:SystemsSchematic}(b)]. Metasurfaces and other periodic structures (gratings, photonic crystal membranes, frequency selective surfaces) are very important components in photonics. In order to be able to tackle the metasurface under study with the proposed framework, we will need to make several modifications; they will be detailed in what follows. 
    
    The length of the graphene strip remains the same, while the periodicity (lattice constant) is chosen as $\Lambda = 10~\mathrm{\mu m}$ (filling factor $f = 50\%$) and the height of the substrate as $h = 19.5~\mathrm{\mu m}$. These two parameters were fine-tuned to achieve perfect absorption for normal incidence with the $m = 3$ order QNM, eliminating any reflections and boosting the conversion efficiency \cite{Theodosi:2021}. 
    In terms of the corresponding quality factors, this is achieved when $Q_\mathrm{res}=Q_\mathrm{rad}$, frequently termed the critical coupling condition \cite{Tsilipakos:2021apl}.
    Furthermore, due to the small wavelength (large parallel wavevector) of propagating plasmons, the metasurface lattice constant is deeply subwavelength. It remains  subwavelength even at the vicinity of the third harmonic frequency ($\lambda_{0,3\omega}\approx 13.4~\mu$m at 22.4~THz), meaning that only the zeroth diffraction order contributes to the reflection at normal incidence. The first diffraction order in reflection will start becoming propagating for $\vartheta \geq 20^\mathrm{o}$, as $\Lambda/\lambda_{0,3\omega} > 1/(\sin{\vartheta} + 1)$, but even for higher angles (e.g. 40$^\mathrm{o}$) considered in the simulations included here, we have found that it carries only a very small portion of the reflected power.
    
    \begin{figure}[t]
        \centering
        \includegraphics{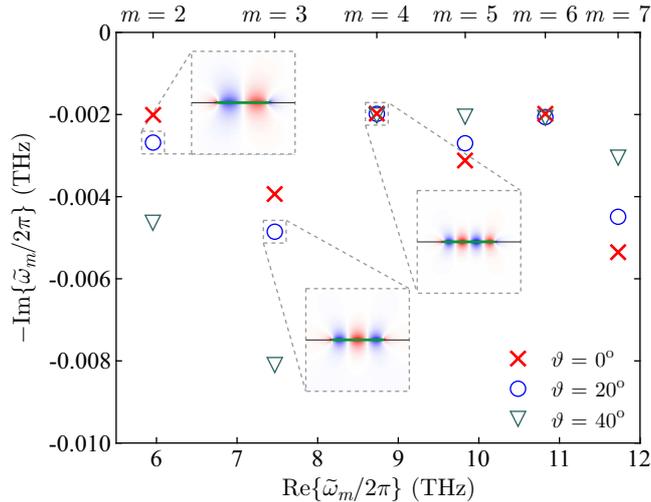}
        \caption{Eigenmode structure of the reflective graphene-strip  metasurface. Dependence of a few, low-order QNMs on the angle of incidence $\vartheta$. The length of the graphene strip is $L=5~\mathrm{\mu m}$, while the pitch of the metasurface is $\Lambda = 10~\mathrm{\mu m}$ and the substrate height is $h=19.5~\mathrm{\mu m}$.
        For details on how the imaginary and real part of the resonance frequencies varies, see the main text and Fig.~\ref{Fig:MetaGrQNMsFull}. Insets: $E_x$-field distribution of a few QNMs for $\vartheta = 20^\mathrm{o}$. 
        }
        \label{Fig:MetaGrQNMs}
    \end{figure}
    
    A small set of the supported QNMs around the fundamental frequency are shown in Fig.~\ref{Fig:MetaGrQNMs}, considering different incidence angles ($\vartheta=0^\mathrm{o}$, $20^\mathrm{o}$, and $40^\mathrm{o}$). They have been calculated assuming the phase advance conditions that would be imposed between the side periodic boundary conditions [planes $x=-\Lambda/2$ and $x=+\Lambda/2$ in  Fig.~\ref{Fig:SystemsSchematic}(b)] in excitation scenarios with obliquely-incident plane waves. Note that the eigenvalue calculations are performed using a formulation involving the periodic envelope of the electric field \cite{Davanco:2007,Fietz:2011} (Bloch-Floquet theorem); the information regarding the phase difference $|\mathbf{k}_F| \Lambda$ between periodic planes, which contains the unknown resonant frequency through $\mathbf{k}_F$, is included within the modified wave equation (instead of a boundary condition) \cite{Gras:2019}.
    See the App.~\ref{subsec:LinearExavulationMeta} for the formulation and a brief discussion. 
    
    Observing Fig.~\ref{Fig:MetaGrQNMs}, the complex eigenfrequency of each mode changes depending on the angle of incidence. Both real and imaginary parts are affected (modifications to the real part are not practically observable due to scaling but are included in Fig.~\ref{Fig:MetaGrQNMsFull}). Interestingly, symmetric ($m$ is odd) and antisymmetric ($m$ is even) modes are affected differently (the symmetry of the modes is considered with respect to the $x=0$ plane). Symmetric modes exhibit pronounced differences in the quality factor, which can either increase or decrease depending on the relation between the resonance wavelength and the height of the substrate. This is verified by considering the case of a transmissive metasurface without a backplane (infinite substrate); in this case, the quality factor obtained form the eigenvalue problem monotonically increases with increasing angle as radiation is suppressed (see App.~\ref{subsec:QNMsSpectumSubstrate}). 
    Antisymmetric modes retain a practically constant quality factor, determined by the resistive component, $Q_\mathrm{res}$, which remains practically unchanged with $\vartheta$ due to the strong confinement of the field and the fact that $Q_\mathrm{rad}$ attains very high values. The only exception is the lowest order mode, $m=2$, which can become quite radiative as the incidence angle increases, falling below $Q_\mathrm{res}$. 
    A more comprehensive discussion regarding the dependence of the QNM eigenvalues on $\vartheta$ is included in App.~\ref{subsec:QNMsSpectumQreconstruction}.
    
    Subsequently, we use the obtained QNMs, as well as the accompanying spurious modes, to calculate the absorption and reflection at the third harmonic frequency. To do so, we have to modify the framework of Sec.~\ref{sec:Framework} to correctly take into account the envelope formulation used to obtain the respective QNMs. To this end, we use the Bloch envelope $\boldsymbol{\tilde\psi}_m$ (lowercase), rather than the full field $\mathbf{\tilde\Psi}_m$, for obtaining the expansion coefficients \cite{Gras:2019}. Using $\mathbf{\tilde\Psi}_m(\mathbf{r}) = \boldsymbol{\tilde\psi}_m(\mathbf{r}) \exp\{-jk_{0,m}\boldsymbol{\eta}\cdot\mathbf{r}\}$ where $\mathbf{k}_{F,m} = k_{0,m}\boldsymbol{\eta}$ is the Bloch wavevector, the curl Maxwell's equations are transformed into $\hat{\mathcal{L}}\boldsymbol{\tilde\psi}_m = \tilde\omega_m\hat{\mathcal{M}}\boldsymbol{\tilde\psi}_m$ and the matrix operator
    \begin{equation}
        \hat{\mathcal{M}} = \begin{bmatrix}  1                                            & -(c_0\mu)^{-1}\boldsymbol{\eta}\times & 0 \\ 
                                             (c_0\varepsilon)^{-1}\boldsymbol{\eta}\times & 1                                     & 0 \\
                                             0                                            & 0                                     & 1
                            \end{bmatrix}, \label{Eq:MLinear}
    \end{equation}
    is introduced to include the information of the incident wave direction through $\boldsymbol{\eta}$. For example, in a metasurface with periodicity along the $x$ axis and under illumination with an incident angle $\vartheta$,  $\boldsymbol{\eta} = n_i \sin\vartheta \mathbf{\hat x}$. Ultimately, the expansion coefficients take the form
    \begin{equation}
        a_m(\omega) = \frac{1}{\tilde\omega_m - \omega}\int_V \mathbf{\tilde j}_{-m} \cdot \mathbf{e}_\mathrm{inc} \mathrm{d}V, \label{Eq:alphaLinearPeriodic}
    \end{equation}
    and the normalization condition acquires an additional term related with $\boldsymbol{\eta}$, now becoming
    \begin{equation}
        \int_V \left[ \mathbf{\tilde e}_{-m}\cdot\varepsilon\mathbf{\tilde e}_m - \mathbf{\tilde h}_{-m}\cdot\mu\mathbf{\tilde h}_m - \mathbf{\tilde e}_{-m}\cdot j\frac{\partial\dbar{\sigma_s}(\omega)}{\partial\omega} \mathbf{\tilde e}_m\delta_s - \boldsymbol{\eta}\cdot\frac{1}{c_0}\left( \mathbf{\tilde h}_{-m}\times\mathbf{\tilde e}_m + \mathbf{\tilde e}_{-m}\times\mathbf{\tilde h}_m \right) \right] \mathrm{d}V = 1. \label{Eq:QNMsPeriodicNormalization}
    \end{equation}
    Note that in Eqs.~\eqref{Eq:alphaLinearPeriodic}~and~\eqref{Eq:QNMsPeriodicNormalization} the notation $\boldsymbol{\tilde \psi}_{-m}$ implies the use of the left eigenvectors \cite{Lalanne:2019,Gras:2019}. In periodic systems comprising reciprocal materials they are found through $\mathbf{\tilde\Psi}_{-m}(\mathbf{r}) = \boldsymbol{\tilde\psi}_{-m}(\mathbf{r}) \exp\{+jk_{0,m}\boldsymbol{\eta}\cdot\mathbf{r}\}$ and $\boldsymbol{\tilde\psi}_{-m}(\mathbf{r})$ can be obtained from $\boldsymbol{\tilde\psi}_{m}(\mathbf{r})$ through simple transformations \cite{Lalanne:2019}. In general, when non-reciprocal materials are involved or if structural asymmetries exist, left eigenvectors need to be calculated separately in order to construct the orthonormal basis that is required to obtain the expansion in QNMs \cite{Kristensen:2020,Weiss:2017,Zhang:2020,Weiss:2018}.
    We highlight again here that $\mathbf{E}_\mathrm{inc}$ should be replaced with $\mathbf{E}_b$ in the presence of a substrate.
    Finally, the expansion coefficients at the third harmonic frequency are calculated through
    \begin{equation}
        a_m(3\omega) = \frac{-j}{\tilde\omega_m - 3\omega}\int_V \left[ \mathbf{\tilde e}_{-m} \cdot \mathbf{j}_{s,\mathrm{NL}}^{(3\omega)}\delta_s - \boldsymbol{\eta}\cdot\displaystyle\frac{1}{c_0\varepsilon}(\mathbf{\tilde h}_{-m}\times\mathbf{j}_{s,\mathrm{NL}}^{(3\omega)}\delta_s) \right] \mathrm{d}V, \label{Eq:alphaNonlinearPeriodic}
    \end{equation}
    and involve an additional term compared to Eq.~\eqref{Eq:alphaNonlinear} which depends on $\boldsymbol{\eta}$. 
    Note that the nonlinear surface current density in $3\omega$ is defined as $\mathbf{j}_{s,\mathrm{NL}}^{(3\omega)} = (\sigma_3/4)  (\mathbf{e}_{t,\parallel}^{(\omega)} \cdot \mathbf{e}_{t,\parallel}^{(\omega)})\mathbf{e}_{t,\parallel}^{(\omega)}$, i.e., naturally by using the envelope $\mathbf{e}$ and $\mathbf{e}_t$ is extracted as the envelope of the total field $\mathbf{E}_t$ by removing the appropriate phase term $\exp\{-jk_0\boldsymbol{\eta}\cdot\mathbf{r}\}$.
    
\begin{figure}[!t]
        \centering
        \includegraphics{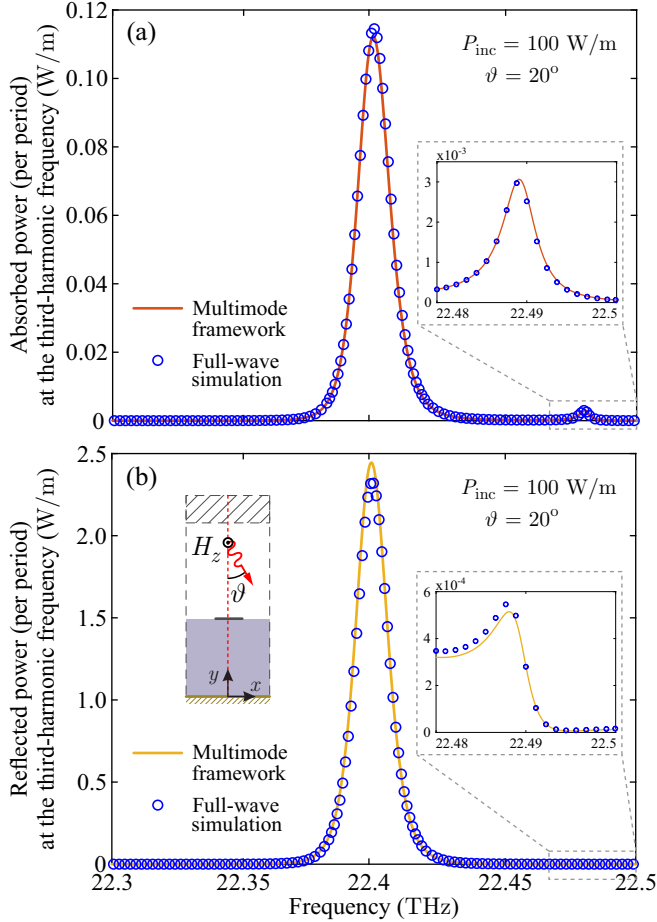}
        \caption{Third-harmonic generation with reflective graphene-strip metasurface. Comparison between the proposed multimode nonlinear framework and full-wave nonlinear simulations. (a) Absorption and (b) reflection at the third harmonic frequency. The incident field is a TM-polarized plane wave ($\mathbf{H}_\mathrm{inc} \equiv H_z\mathbf{\hat z}$) at an incidence angle $\vartheta = 20^\mathrm{o}$. Insets: Zoom-in around $22.49~\mathrm{THz}$ where the second peak lies. The capabilities of the framework in reproducing the full-wave results are evident, even when weak and non-Lorentzian lineshapes are involved.}
        \label{Fig:MetaGrNLAbsRefl}
    \end{figure}

    Next, the structure is excited with a TM-polarized plane wave inside the $xy$-plane ($\mathbf{H}_\mathrm{inc} \equiv H_z\mathbf{\hat z}$) and the obtained results (reflected and absorbed power) are depicted in Fig.~\ref{Fig:MetaGrNLAbsRefl} for an input power (per period) of $100~\mathrm{W/m}$ and an angle of incidence  $\vartheta = 20^\mathrm{o}$. 
    Note that this input power level corresponds to an electric field of $|E_\mathrm{inc}^{(\omega)}| = 0.86~\mathrm{kV/cm}$. This value is within current experimental capabilities. For example, in Ref.~\cite{Hafez:2019AdvOptMat} an experiment was conducted at 0.3 THz with a peak electric field (pulsed operation) of $|E_\mathrm{inc}^{(\omega)}| \sim 80~\mathrm{kV/cm}$ to reveal up to seventh-harmonic generation. For the chosen input power level though, only THG is practically significant.
    Once again, excellent agreement with full-wave nonlinear simulations is attained. The main peak appears at $22.404~\mathrm{THz}$, exactly at $3\omega_\mathrm{res,3}$, while a second peak appears in this example as well, lying at $\omega_\mathrm{res,25}/2\pi = 22.489~\mathrm{THz}$, which is the resonance frequency of the respective $m=25$ QNM. In contrast to Fig.~\ref{Fig:SingleGrNLXsections}(a), the amplitude of this secondary peak is significantly suppressed since the fundamental frequency $22.489/3 = 7.496~\mathrm{THz}$ is not accommodated under the respective Lorentzian, which is centered at $\omega_\mathrm{res,3}/2\pi = 7.468~\mathrm{THz}$ and possesses $Q_{i,3} = 769.3$. Nevertheless, the proposed framework succeeds in accurately capturing the amplitude and lineshape of these weaker peaks as well (see insets of Fig.~\ref{Fig:MetaGrNLAbsRefl}).
    
\begin{figure}[!t]
        \centering
        \includegraphics{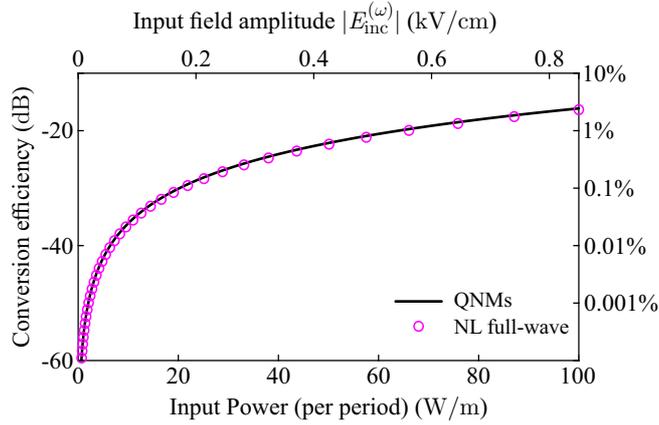}
        \caption{Conversion efficiency of the graphene-strip metasurface versus incident power. The results obtained using the proposed QNM framework are in excellent agreement with nonlinear full-wave simulations over a wide range of $P_\mathrm{inc}$ values.
        The conversion efficiency can exceed $-20~\mathrm{dB}$ (1\%) for realistic input intensities (see main text); this efficient up-conversion is a result of the strong nonlinearity of graphene and the high quality factor of the supported resonances.
        }
        \label{Fig:MetaGrNLCE}
    \end{figure}

    Finally, in Fig.~\ref{Fig:MetaGrNLCE} we keep the incident wave frequency constant at $7.468~\mathrm{THz}$, vary the power it carries and examine the conversion efficiency from the fundamental to the third harmonic frequency [calculated as $\mathrm{CE} = 10\log(P_\mathrm{refl}^{(3\omega)}/P_\mathrm{inc})$]. The conversion efficiency increases with incident power as anticipated and exceeds $-20~\mathrm{dB}$ (1\%); this efficient up-conversion is a result of the strong nonlinearity of graphene and the high quality factor of the supported resonances.
    Clearly, the proposed framework and the nonlinear full-wave simulations are in excellent agreement over a wide range of $P_\mathrm{inc}$ values. Note here that the developed framework and the full-wave simulations are performed for the undepleted-pump scenario, i.e., assuming that a small portion of the pump power is converted to the third harmonic frequency and additionally that back-conversion from the third harmonic to the fundamental frequency is negligible. These assumptions are very reasonable for CEs up to $-20~\mathrm{dB}$ (1\%) \cite{Christopoulos:2018}.
    
\section{\label{sec:Conclusions} Conclusion}

    We have proposed a modal formalism for studying third harmonic generation in non-Hermitian open resonant systems comprising 2D materials. It is constructed by expressing the scattered fields at the fundamental and third harmonic frequencies as expansions in a set of supported quasinormal modes. The infinitesimally-thin and dispersive nature of 2D materials has been rigorously taken into account. A graphene-strip single scatterer and a periodic metasurface were selected as examples for demonstrating the framework capabilities. In both cases, excellent agreement with full-wave nonlinear simulations has been obtained. By providing insight into the full range of quasinormal modes supported by the structures, we were able to explain the spectral features of the nonlinear response arising due to the different resonant modes involved in the conversion process. In the reflective metasurface, the conversion efficiency can exceed $-20~\mathrm{dB}$ (1\%) for realistic input intensities, highlighting the practical potential of graphene and 2D materials in general for nonlinear nanophotonics. The proposed framework combines computational efficiency along with the ability to acquire physical insight into the frequency generation process and provide guidelines for boosting the conversion by engineering the mode spectrum.
    
\section*{Acknowledgements}
    This work was supported by the Hellenic Foundation for Research and Innovation (H.F.R.I.) under the ``2nd Call for H.F.R.I. Research Projects to support Post-doctoral Researchers'' (Project Number: 916, PHOTOSURF)

\appendix
\numberwithin{equation}{section}
\numberwithin{figure}{section}

\section*{Appendices}
    The following Appendices contain the complete derivation of the proposed framework,  additional simulation results to support the discussions included in the main manuscript, and provides deeper  physical insight into the graphene-based resonant systems under study. Specifically, it contains: A general version of the framework which includes lossy, dispersive, and anisotropic materials; more detailed documentation of the supported modes in the graphene-strip scatterer and graphene-strip metasurface; validation of the linear version of the framework; evidence regarding the necessity of the surface term in the normalization of the QNMs; more details regarding the $\mathbf{e}$-field envelope formulation.

\section{\label{sec:GeneralFramework} Multimode quasinormal-mode framework for photonic scatterers incorporating bulk and sheet third-order nonlinear materials}

    In this first section, we present the complete derivation of our developed framework, which includes the general case of cavities with lossy, dispersive, and anisotropic materials. Furthermore, the resonant cavity can include both bulk and sheet-type materials. Although any combination is allowed, for the sake of the presentation we will assume a cavity consisting of a bulk material with a single Drude-Lorentz pole and a sheet material with a single Drude pole, i.e., described by the equations
    \begin{subequations}
        \begin{align}
            \dbar{\varepsilon}(\omega) &= \varepsilon_0\dbar{\varepsilon}_\infty \left( 1 - \frac{\omega_p^2}{\omega^2 - \omega_0^2 - j\omega\gamma_b} \right), \\
            \dbar{\sigma}_s(\omega) &= -j\frac{\dbar{\sigma}_0}{\omega - j\gamma_s}.
        \end{align}
        \label{Eq:Matrials}
    \end{subequations}
    The parameters $\omega_p$, $\omega_0$, and $\gamma_b$ are, respectively, the plasma frequency, resonance frequency, and damping factor of the bulk material. Additionally, $\dbar{\varepsilon}_\infty$ is the relative permittivity of the material at infinite frequency, which can be anisotropic. For the sheet material, $\gamma_s$ is the damping factor and $\dbar{\sigma}_0/\gamma_s$ is the (anisotropic) conductivity at $\omega = 0$. Note that both $\dbar{\varepsilon}$ and $\dbar{\sigma}_s$ are spatially dependent, so that $\dbar{\sigma}_s(\mathbf{r}) = \mathbf{0}$ outside the 2D material. For bulk materials without dispersion, as for example is commonly the case with the background permittivity $\varepsilon_b$ of a photonic scatterer, $\dbar{\varepsilon}(\mathbf{r}) = \varepsilon_0\varepsilon_b\mathbf{I}_3$ ($\mathbf{I}_3$ is the $3\times3$ identity matrix).
    
    To include both the bulk and sheet dispersive material in the framework, three auxiliary fields are needed, two for the Drude-Lorentz pole of the bulk material and one for the Drude pole of the 2D material \cite{Yan:2018,Raman:2011}. More generally, a Drude-Lorentz pole requires always two auxiliary fields and a Drude pole only one. In our case, we define the auxiliary fields
    \begin{subequations}
        \begin{align}
            \mathbf{P}_b &= -\varepsilon_0\dbar{\varepsilon}_\infty \frac{\omega_p^2}{\omega^2 - \omega_0^2 - j\omega\gamma_b} \mathbf{E}, \\
            \mathbf{J}_b &= j\omega \mathbf{P}_b, \\
            \mathbf{J}_s &= -\frac{\dbar{\sigma}_0}{\omega - j\gamma_s}\mathbf{E}\delta_s,
        \end{align}
        \label{Eq:AuxFields}
    \end{subequations}
    to accompany the electric $\mathbf{E}$ and magnetic $\mathbf{H}$ fields in the Maxwell's equations. Following the notation of the manuscript, the source-free curl Maxwell's equations are expressed in the compact notation $\hat{\mathcal{L}}\mathbf{\tilde\Psi}_m = \tilde\omega_m\mathbf{\tilde\Psi}_m$, where now $\mathbf{\tilde\Psi}_m = [\mathbf{\tilde H}_m~~\mathbf{\tilde E}_m~~\mathbf{\tilde P}_{b,m}~~\mathbf{\tilde J}_{b,m}~~\mathbf{\tilde J}_{s,m}]^T$ and
    \begin{equation}
        \hat{\mathcal{L}} = \begin{bmatrix}  0 & j\mu^{-1}\nabla\times & 0 & 0 & 0 \\ 
                                             -j(\varepsilon_0\dbar{\varepsilon}_\infty)^{-1}\nabla\times & 0 & 0 & j(\varepsilon_0\dbar{\varepsilon}_\infty)^{-1} & -(\varepsilon_0\dbar{\varepsilon}_\infty)^{-1} \\
                                             0 & 0 & 0 & -j & 0 \\
                                             0 & -j(\varepsilon_0\dbar{\varepsilon}_\infty)\omega_p^2 & j\omega_0^2 & j\gamma_b & 0 \\
                                             0 & -\dbar{\sigma}_0\delta_s & 0 & 0 & j\gamma_s
                            \end{bmatrix}.
        \label{Eq:LLinearSuppl}
    \end{equation}

    To find the expansion coefficients $a_m(\omega)$, one has to begin from the scattering formulation at the fundamental frequency. Then, the curl Maxwell's equations take the compact form $\hat{\mathcal{L}}\mathbf{\Psi}_\mathrm{sct}^{(\omega)} = \omega\mathbf{\Psi}_\mathrm{sct}^{(\omega)} + \mathbf{S}_\mathrm{inc}^{(\omega)}$, where 
    \begin{equation}
        \mathbf{S}_\mathrm{inc}^{(\omega)} = \begin{bmatrix} \mathbf{0} \\
                                                             \omega(1 - \varepsilon_b\dbar{\varepsilon}_\infty^{-1})\mathbf{E}_\mathrm{inc} \\
                                                             \mathbf{0} \\
                                                             j\omega_p^2\varepsilon_0\dbar{\varepsilon}_\infty\mathbf{E}_\mathrm{inc} \\ 
                                                             \dbar{\sigma}_0\mathbf{E}_\mathrm{inc}\delta_s
                                            \end{bmatrix},
        \label{Eq:SincLinFull}
    \end{equation}
    showing that the incident field interacts both with the bulk and sheet material. Using Eq.~\eqref{Eq:SincLinFull} it is easy to show that the expansion coefficients $a_m(\omega)$ are given by the general expression \cite{Yan:2018}
    \begin{equation}
            a_m(\omega) = \frac{1}{\tilde\omega_m - \omega} \frac{\displaystyle\int_V \mathbf{\tilde\Psi}_m^T\mathbf{\hat D}\mathbf{S}_\mathrm{inc}^{(\omega)}\mathrm{d}V}{\displaystyle\int_V \mathbf{\tilde\Psi}_m^T\mathbf{\hat D}\mathbf{\tilde\Psi}_m\mathrm{d}V},
        \label{Eq:amLinFull}
    \end{equation}
    with $\mathbf{\hat D} = \mathrm{diag}(-\mu,~ \varepsilon_0\dbar{\varepsilon}_\infty,~ (\omega_p^2\varepsilon_0\dbar{\varepsilon}_\infty)^{-1}\omega_0^2,~ -(\omega_p^2\varepsilon_0\dbar{\varepsilon}_\infty)^{-1},~ \dbar{\sigma}_0^{-1})$ being a diagonal matrix, suitable to apply the unconjugated Lorentz reciprocity theorem \cite{Raman:2011}. The denominator of Eq.~\eqref{Eq:amLinFull} is actually the QNM normalization factor, which can be cast in the expanded form
    \begin{equation}
        \int_V \left( \mathbf{\tilde E}_m\cdot\left.\frac{\partial\{\omega\dbar{\varepsilon}(\omega)\}}{\partial\omega}\right|_{\omega = \tilde\omega_m} \mathbf{\tilde E}_m - \mathbf{\tilde H}_m\cdot\mu\mathbf{\tilde H}_m - \mathbf{\tilde E}_m\cdot j\left.\frac{\partial\dbar{\sigma}(\omega)}{\partial\omega}\right|_{\omega = \tilde\omega_m}\mathbf{\tilde E}_m\delta_s \right) \mathrm{d}V = 1,
        \label{Eq:QNMsNormalizationFull}
    \end{equation}
    and includes the dispersive properties of both bulk and sheet-type materials through the spectral derivatives $\partial\{\omega\dbar{\varepsilon}(\omega)\}/\partial\omega$ and $\partial\dbar{\sigma}(\omega)/\partial\omega$, respectively. Henceforth, we assume that all the QNMs are appropriately normalized according to  Eq.~\eqref{Eq:QNMsNormalizationFull}. With this in mind, the expansion coefficients of Eq.~\eqref{Eq:amLinFull} take the more useful form
    \begin{align}
        a_m(\omega) = &\frac{1}{\tilde\omega_m - \omega}\times \nonumber\\
                      &\int_V\left[ \omega\varepsilon_0(\dbar{\varepsilon}_\infty - \varepsilon_b)\mathbf{\tilde E}_m \cdot \mathbf{E}_\mathrm{inc} + \tilde\omega_m\varepsilon_0(\dbar{\varepsilon}_r(\tilde\omega_m) - \dbar{\varepsilon}_\infty)\mathbf{\tilde E}_m \cdot \mathbf{E}_\mathrm{inc} - \frac{\dbar{\sigma}_0}{\tilde\omega_m-j\gamma_s}\mathbf{\tilde E}_m\delta_s \cdot \mathbf{E}_\mathrm{inc} \right]\mathrm{d}V.
        \label{Eq:alphaLinearFull}
    \end{align}
    It is easy to see how Eq.~\eqref{Eq:alphaLinearFull} expands the framework of Ref.~\cite{Yan:2018}, since it now includes the contribution of both bulk and sheet-type, dispersive materials. One can also see that Eq.~\eqref{Eq:alphaLinear} is a simplified version of Eq.~\eqref{Eq:alphaLinearFull}.
    Note that in Eq.~\eqref{Eq:alphaLinearFull}, we have used the notation $\dbar{\varepsilon}_r(\tilde\omega_m) = \dbar{\varepsilon}(\tilde\omega_m)/\varepsilon_0$. Furthermore, we shall highlight again that in the presence of a reflected wave, the incident field $\mathbf{E}_\mathrm{inc}$ should be replaced with the background field $\mathbf{E}_b$ throughout the calculations.
    
    At the third harmonic frequency, a similar approach can be followed to retrieve the respective coefficients. We allow for the general case where both the bulk and sheet-type material exhibit nonlinearities of the same form, induced using the nonlinear counterpart of the auxiliary fields $\mathbf{P}_b$ and $\mathbf{J}_s$, respectively, i.e., $\mathbf{P}_{b,\mathrm{NL}}^{(3\omega)} = (\chi_3/4)  (\mathbf{E}_t^{(\omega)} \cdot \mathbf{E}_t^{(\omega)})\mathbf{E}_t^{(\omega)}$ and $\mathbf{J}_{s,\mathrm{NL}}^{(3\omega)} = (\sigma_3/4)  (\mathbf{E}_{t,\parallel}^{(\omega)} \cdot \mathbf{E}_{t,\parallel}^{(\omega)})\mathbf{E}_{t,\parallel}^{(\omega)}$, with the subscript ``$t$'' denoting the total field as the sum of the incident, reflected from the background, and scattered fields. The use of these auxiliary fields which actually coincide with the respective nonlinear polarization and surface current quantities is a natural selection when electronic nonlinearities are described \cite{Christopoulos:2018}. The scattered field formulation at $3\omega$ acquires the compact form $\hat{\mathcal{L}}\mathbf{\Psi}_\mathrm{sct}^{(3\omega)} = 3\omega\mathbf{\Psi}_\mathrm{sct}^{(3\omega)} + \mathbf{S}_\mathrm{inc}^{(3\omega)}$, where the source term is given by
    \begin{equation}
        \mathbf{S}_\mathrm{inc}^{(3\omega)} = \begin{bmatrix} \mathbf{0} \\
                                                             -j(\varepsilon_0\dbar{\varepsilon}_\infty)^{-1}[ j(3\omega)\mathbf{P}_{b,\mathrm{NL}} + \mathbf{J}_{s,\mathrm{NL}}\delta_s ] \\
                                                             \mathbf{0} \\
                                                             \mathbf{0} \\
                                                             \mathbf{0}
                                            \end{bmatrix},
        \label{Eq:Eq:SincNonLin}
    \end{equation}
    i.e., it only contributes as a right-hand-side term in the Amp\`ere-Maxwell equation. Using $\mathbf{S}_\mathrm{inc}^{(3\omega)}$ in Eq.~\eqref{Eq:amLinFull}, the expansion coefficients for the third harmonic frequency can be calculated through
    \begin{equation}
        a_m(3\omega) = \frac{1}{\tilde\omega_m - 3\omega}\int_V\left( 3\omega\mathbf{\tilde E}_m \cdot \mathbf{P}_{b,\mathrm{NL}}^{(3\omega)} - j\mathbf{\tilde E}_m \cdot \mathbf{J}_{s,\mathrm{NL}}^{(3\omega)}\delta_s \right)\mathrm{d}V.
        \label{Eq:alphaNonLinFull}
    \end{equation}
    Equation~\eqref{Eq:alphaNonLinFull} is a generalized version of Eq.~\eqref{Eq:alphaNonlinear} to include the more general case of cavities with both bulk and sheet-type third-order nonlinear materials. To the best of our knowledge, this is the first time such an equation is presented in the context of \emph{multimode} \emph{non-Hermitian} systems. In our recent works, the simpler case of \emph{single-mode} \emph{quasi-Hermitian} systems has been addressed \cite{Christopoulos:2016PRE,Christopoulos:2018}.
    
    As a final remark, we note that materials with multiple Drude-Lorentz poles can be introduced in the formulation simply by using an appropriate number of auxiliary fields to represent the respective poles. Examples of such a scenario are noble metals with interband transitions above the plasma frequency or dielectric materials described by a Sellmeier equation of the general form
    \begin{equation}
        n^2(\lambda) = n_0^2 + \sum_{k=1}^3 \frac{B_k\lambda^2}{\lambda^2-C_k^2}.
        \label{Eq:Sellmeier}
    \end{equation}
    Equation~\eqref{Eq:Sellmeier} introduces three poles in the system, thus six auxiliary fields are needed for its correct representation. It is more convenient to express Eq.~\eqref{Eq:Sellmeier} as a function of the angular frequency instead of the wavelength and to use the permittivity instead of the refractive index, i.e.,
    \begin{equation}
        \varepsilon(\omega) = \varepsilon_0\varepsilon_\infty\left(1 - \sum_{k=1}^3 \frac{\omega_{p,k}^2}{\omega^2-\omega_{0,k}^2}\right),
        \label{Eq:SellmeierFreq}
    \end{equation}
    where $\omega_{p,k} = 2\pi c_0\sqrt{B_k/(n_0 C_k)^2}$ and $\omega_{0,k} = 2\pi c_0\sqrt{1/C_k^2}$ are the plasma frequency and the resonance frequency of each pole while $\varepsilon_\infty \equiv n_0^2$; typically no losses are included in the Sellmeier equation since it is used to describe dielectric materials and thus $\gamma_{p,k} = 0$. 
    The expansion of the framework to resonant structures consisting of multiple bulk/sheet materials with multiple poles each is trivial following the above description.

\section{\label{sec:ExamplesSingle} Graphene-strip single scatterer on a glass substrate}

    In this section, additional simulation results for the single-scatterer example are provided. The results further support the framework presented in the manuscript and include: (i) The full spectrum of QNM and spurious modes used for the calculations, (ii) an evaluation of the linear framework and a comparison with another well-established analytic theory, the temporal coupled-mode theory (CMT), and (iii) a proof of the necessity of the extra term in the QNM normalization when dispersive sheet materials are included.
    
\subsection{\label{subsec:QNMsSpectumFull} Full spectrum of QNMs and spurious modes}

\begin{figure}[t]
        \centering
        \includegraphics{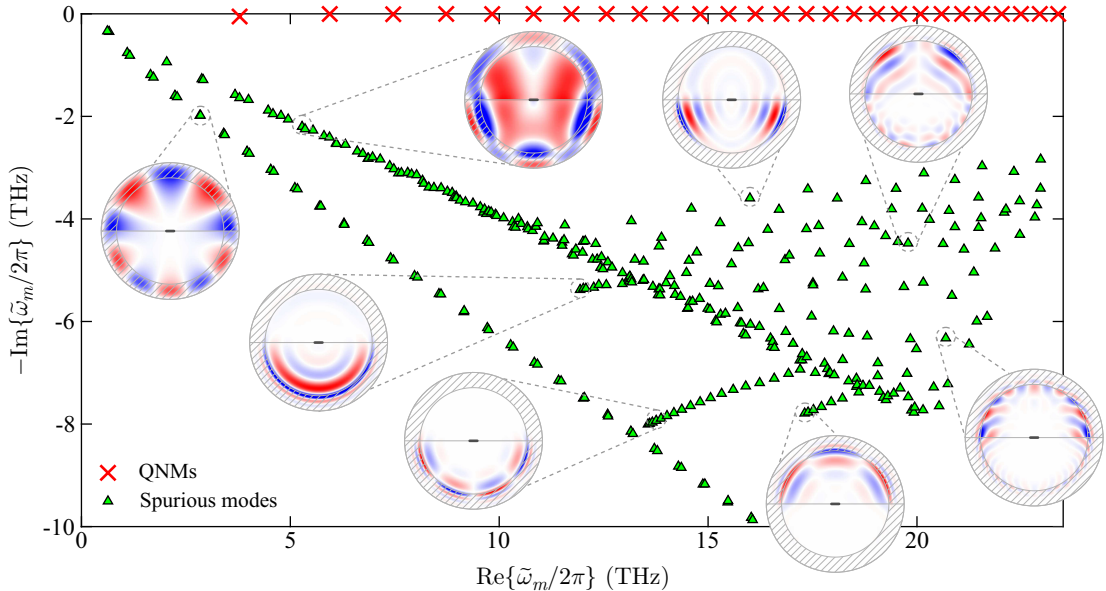}
        \caption{Extended spectrum of modes (approximately 300 are depicted) of the examined graphene-strip scatterer. The modes are categorized as either QNMs (red X marks) or spurious modes (green triangle marks). The PML-induced rotation of the modes' spectrum \cite{Sauvan:2022} is clearly seen. Other non-physical modes are found and included as well. Inset: field distribution of a few spurious modes. They exhibit a rich variety in terms of their field distribution; in some cases the field is mostly located inside the PML region (PML modes), while in others the field distribution occupies the whole computational domain (other non-physical modes).}
        \label{Fig:SingleGrQNMsFull}
    \end{figure}

    In Fig.~\ref{Fig:SingleGrQNMsFull}, the full set of the eigenmodes used for the calculations (approximately 300) is presented. It includes QNMs, clearly marked with red ``X'' indicators, and spurious (i.e., non-physical) modes, indicated with green triangles. Note that for this specific example, such a huge number of modes is not strictly necessary since each QNM is well-separated from its neighbouring QNMs (due to their Fabry-P\'erot nature), and with a relatively high quality factor. Nevertheless, in a more complex structure with a denser QNM distribution (and/or with lower quality factors), such a large number of modes will be necessary \cite{Yan:2018}.
    
    As an indication of the variety of field distributions that spurious modes exhibit, in Fig.~\ref{Fig:SingleGrQNMsFull} we include some as insets. The rotation of the spectrum caused by the PML's complex coordinate transformation \cite{Sauvan:2022} is clearly seen, and is further enriched due to the non-uniformity of the computational space (glass substrate). Such modes are sometimes referred to as PML modes. All these modes are considered to hold information related with the transformation of the computational space from infinite to finite, i.e., computational domain truncation. In addition, a large number of other non-physical modes are found, with their field distribution mostly located outside the PML region. Such modes are a result of the discretization and the finite number of basis functions used for the evaluation of the Galerkin method.
    
\begin{figure}[b]
        \centering
        \includegraphics{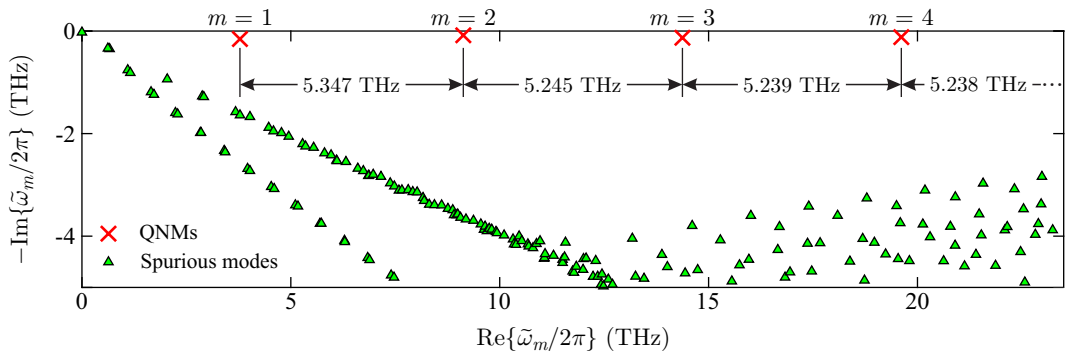}
        \caption{Spectrum of modes for the graphene-strip scatterer when material dispersion is momentarily neglected. A quite smaller number of QNMs are found in the same frequency window. They are almost equidistant, as anticipated due to their Fabry-P\'erot nature.}
        \label{Fig:SingleGrQNMsNonDisp}
    \end{figure}

    In Fig.~\ref{Fig:SingleGrQNMsNonDisp} we depict the pole structure when graphene conductivity is considered frequency-independent and equal to $\dbar{\sigma}_s = \dbar{\sigma}_s(\mathrm{Re}\{\tilde\omega_1\})$. In this nondispersive (unphysical) system, only four QNMs are supported in the same frequency window as the one in Fig.~\ref{Fig:SingleGrQNMsFull} and are furthermore almost equidistant. This is anticipated due to the Fabry-P\'erot nature of the supported modes. The small dependence of the free-spectral range (FSR) on frequency emerges from any dispersion of the reflection phase experienced at the termination of the strip \cite{Nikitin:2014}. 
    Another interesting observation is that almost all spurious modes have the same eigenvalue (complex resonance frequency) with the previous, dispersive case, since they mainly reside in the background materials and the PMLs and not affected by the presence (or absence) of graphene dispersion.
    
\subsection{\label{subsec:LinearExavulationSinge} Evaluation of the linear framework}

\begin{figure}[t]
        \centering
        \includegraphics{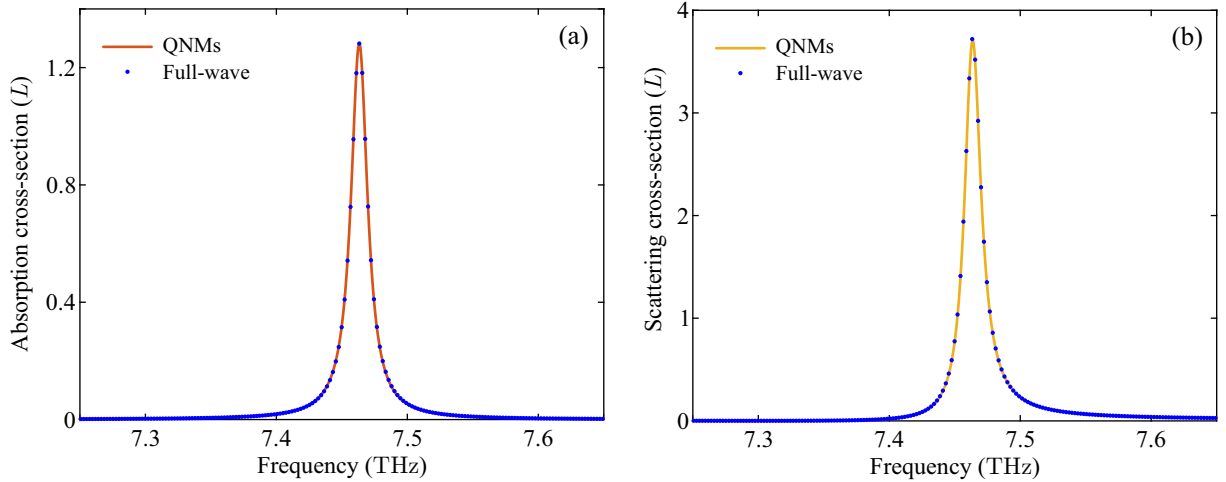}
        \caption{Comparison of the proposed QNMs framework (solid lines) with full-wave simulations (dot markers). (a)~Absorption cross-section and (b)~scattering cross-section. The two methods are in excellent agreement, even when the lineshape is asymmetric (scattering cross-section).}
        \label{Fig:SingleGrLinearComparison}
    \end{figure}

    Using the full spectrum of modes in Fig.~\ref{Fig:SingleGrQNMsFull} (or an appropriately selected smaller set), one is able to reconstruct the linear absorption and scattering cross-sections of the examined graphene strip under a plane-wave illumination of the form $\mathbf{E}_\mathrm{inc} = E_0\exp\{+jk_0y\}\mathbf{\hat x}$, propagating in the $-\mathbf{\hat y}$ direction. First, the expansion coefficients should be calculated through Eq.~\eqref{Eq:alphaLinear}. Then the scattered field on graphene is reconstructed through $\mathbf{E}_\mathrm{sct}(\omega) = \sum_m a_m(\omega) \mathbf{\tilde E}_m$. Finally, the absorption cross-section is easily obtained through a single integration on graphene, i.e.,
    \begin{equation}
        C_\mathrm{abs} = \frac{1}{2I_0} \int_\ell \mathrm{Re}\{\sigma_s(\omega)\}|\mathbf{E}_\mathrm{sct,\parallel}|^2 \mathrm{d}\ell = \frac{1}{2I_0} \int_\ell \frac{\gamma}{\sigma_0} |\mathbf{J}_\mathrm{sct,\parallel}|^2 \mathrm{d}\ell,
        \label{Eq:CabsQNM}
    \end{equation}
    where $I_0=E_0^2/2\eta_0$ is the illumination field intensity, $\eta_0 \simeq 120\pi$ is the free-space impedance, and $\sigma_s(\omega)$ is the only independent element of the $\dbar{\sigma}_s(\omega)$ tensor; for instance $\sigma_s(\omega) = \sigma_{xx}(\omega) = \sigma_{zz}(\omega) = \sigma_0/(\omega - j\gamma)$ for a 2D material laid on the $xz$ plane. For the scattering cross-section, it  is more convenient to  perform the calculation implicitly using the scattered field on graphene, rather than explicitly on a closed curve enclosing the scatterer. To do so, one has to calculate first the extinction cross-section through \cite{Bai:2013}
    \begin{equation}
        C_\mathrm{ext} = -\frac{1}{2I_0} \int_\ell \mathrm{Im}\left\{ \mathbf{J}_\mathrm{sct,\parallel}^*\cdot\mathbf{E}_\mathrm{sct,\parallel} \right\}\mathrm{d}\ell = \frac{1}{2I_0} \int_\ell \mathrm{Im}\left\{ \frac{\sigma_0}{\omega+j\gamma} \mathbf{E}_\mathrm{sct,\parallel}^*\cdot\mathbf{E}_\mathrm{sct,\parallel} \right\}\mathrm{d}\ell,
        \label{Eq:CextQNM}
    \end{equation}
    and then simply subtract the two quantities, i.e., $C_\mathrm{sct} = C_\mathrm{ext} - C_\mathrm{abs}$. The results of Fig.~\ref{Fig:SingleGrLinearComparison} show perfect agreement between the multimode QNM framework and full-wave simulations. Importantly, the asymmetric lineshape in Fig.~\ref{Fig:SingleGrLinearComparison}(b) is accurately reproduced.
    
    Another classical approach to estimate the linear response of the cavity is to use the temporal CMT framework \cite{Hamam:2007}. CMT is a \emph{single-mode} theory developed for Hermitian or quasi-Hermitian systems. To evaluate the response with CMT, the resonance frequency of a single resonance and the respective quality factors (for free-space examples, resistive and radiation quality factors) are needed. These quantities can be obtained through the same eigenvalue simulations (for a discussion regarding the retrieval of the quality factors, see Sec.~\ref{subsec:QNMsSpectumQreconstruction} or Ref.~\cite{Christopoulos:2019OpEx}). The respective equation to calculate absorption and scattering cross-sections are \cite{Hamam:2007}
    \begin{subequations}
        \begin{equation}
            C_\mathrm{abs} = \frac{2(m+1)\lambda}{\pi}\,\frac{r_Q}{\delta^2+(1+r_Q)^2},
            \label{Eq:CabsCMT}
        \end{equation}
        \begin{equation}
            C_\mathrm{sct} = \frac{2(m+1)\lambda}{\pi}\,\frac{1}{\delta^2+(1+r_Q)^2},
            \label{Eq:CextCMT}
        \end{equation}
    \end{subequations}
    where $r_Q = Q_\mathrm{rad}/Q_\mathrm{res}$, $\delta = 2Q_\mathrm{rad}(\omega-\omega_m)/\omega_m$, and $\lambda = \lambda_0/n$. We should note here that this form of CMT is developed for a uniform background and a bulk dielectric resonator. Thus, in our system the choice of $n$ is ambiguous. Furthermore, CMT is capable of reproducing only Lorentzian lineshapes. This precludes cavities with overlapping resonances or the case of  a background wave interacting with the scattered wave, in which case the phase relation between the two should be appropriately accounted for. In such cases, a fitting process is followed in the literature \cite{Fan:2003}.
    Thus, although single-mode temporal CMT is a powerful tool, the study of contemporary non-Hermitian systems requires the  multimode framework developed in this work.
    

\subsection{\label{subsec:QNMsNormalization} Examination of the surface term in the QNMs normalization}

    The last topic we are discussing in the single graphene-strip example is the correct normalization of the QNMs. In the developed framework, a new term that describes the contribution of the 2D material [see Eq.~\eqref{Eq:QNMsNormalization}] is introduced. It is  repeated here using a slightly different notation
    \begin{equation}
        \int_\ell \sigma_0^{-1} \mathbf{\tilde J}_{m,\parallel}\cdot\mathbf{\tilde J}_{m,\parallel} \mathrm{d}\ell = - \int_\ell j\left.\frac{\partial\sigma(\omega)}{\partial\omega}\right|_{\omega = \tilde\omega_m} \mathbf{\tilde E}_{m,\parallel}\cdot\mathbf{\tilde E}_{m,\parallel} \mathrm{d}\ell.
        \label{Eq:QNgr}
    \end{equation}
    It is easily seen that this term acquires non-zero values when the 2D material exhibits dispersion and obtains large values when dispersion is strong, as is the case with graphene and other 2D materials. Note that a similar term has been also obtained in the respective normalization of quasi-Hermitian systems recently \cite{Christopoulos:2016PRE}.

    \begin{figure}[t]
        \centering
        \includegraphics{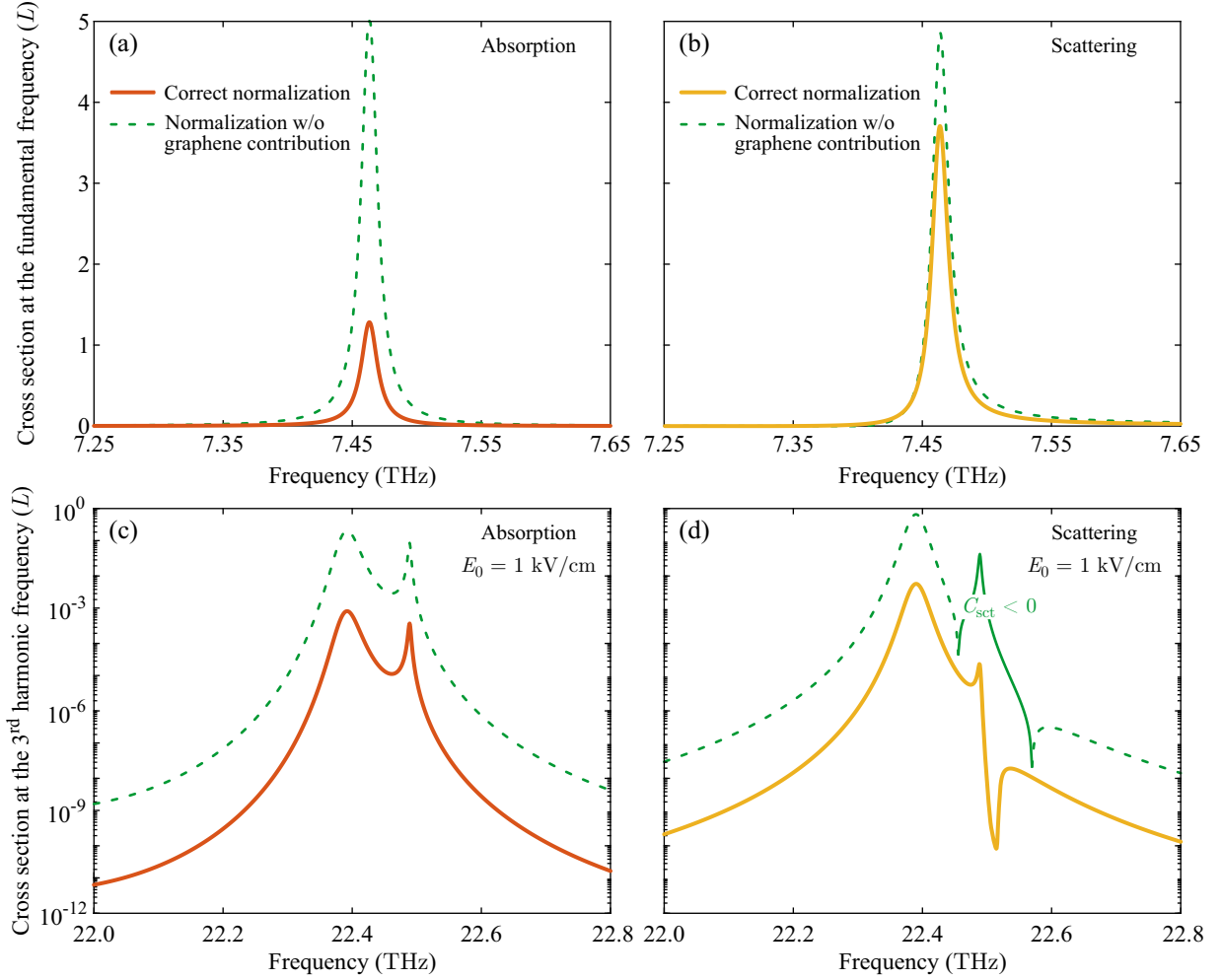}
        \caption{Absorption and scattering cross-sections when including (solid lines) or omitting (dashed lines) the  surface term of Eq.~\eqref{Eq:QNgr} in the normalization. (a-b) Absorption and scattering cross-sections at the fundamental frequency $\omega$ and (c-d) absorption and scattering cross-sections at the third-harmonic frequency $3\omega$ for $E_0 = 1~\mathrm{kV/cm}$. The errors when omitting the surface term in  the normalization are significant and even lead to unphysical results, i.e., negative values for the scattering cross-section denoted with the solid green line in panel (d).}
        \label{Fig:SingleGrNonlinearComparison}
    \end{figure}
    
    In Fig.~\ref{Fig:SingleGrNonlinearComparison}, we compare absorption and scattering cross-sections at the fundamental and third-harmonic frequencies when including (solid lines) or omitting (dashed lines) the term in Eq.~\eqref{Eq:QNgr}. Clearly, the differences are notable both for linear [panels (a-b)] and nonlinear [panels (c-d)] calculations. It becomes obvious that the contribution of this term in the normalization process is significant and should be always included. For the cross-sections at $3\omega$, where the normalized QNMs are considered three times in the $\mathbf{J}_{s,\mathrm{NL}}^{(3\omega)}$ expression, the discrepancy is as large as three orders of magnitude [panels (c) and (d) are depicted in log scale]. Furthermore, the scattering cross-section acquires unphysical negative values [green solid line in Fig.~\ref{Fig:SingleGrNonlinearComparison}(d)], which would mean that power is inserted in rather than scattered by the cavity.

\section{\label{sec:ExamplesMeta} Graphene-strip metasurface on a metal-backed glass substrate}

    In this section, more computational results for the graphene-strip metasurface example are provided: (i) Decomposition into the constituent quality factors (radiative, resistive) to understand the dependencies on the incidence wave angle $\vartheta$, (ii) examination of a transmissive version of the metasurface with an infinite substrate, (iii) evaluation of the linear framework with additional focus on the necessity of the $\mathbf{e}$-field formulation and the respective normalization, and (iv) an additional figure showing the nonlinear absorption of the metasurface.

\subsection{\label{subsec:QNMsSpectumQreconstruction} $Q$-factor decomposition}

\begin{figure}[!t]
        \centering
        \includegraphics{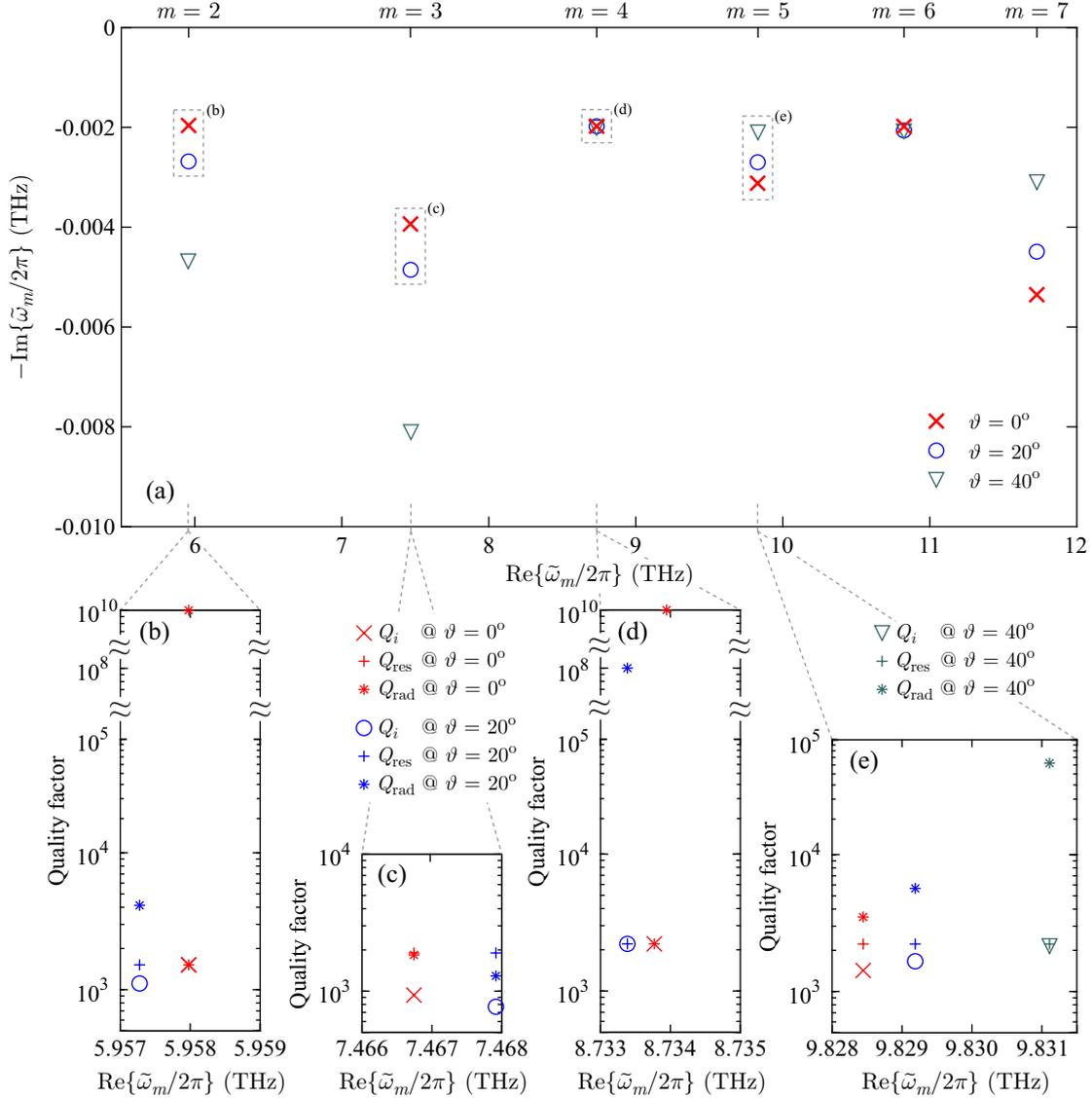}
        \caption{ (a) Dependence of quality factor ($Q_i \propto 1/|\mathrm{Im}\{\tilde\omega_m\}|$) on incidence angle $\vartheta$ for a few, low-order QNMs of the graphene-strip metasurface. (b)-(e)~Decomposition of the intrinsic quality factor into its resistive and radiation constituents for the $m=2$ to $m=5$ order modes.}
        \label{Fig:MetaGrQNMsFull}
    \end{figure}

    To better understand the response of the graphene-strip metasurface under different angles of incidence, in Fig.~\ref{Fig:MetaGrQNMsFull}(a) we plot the real and imaginary parts of a few low-order QNMs. For QNMs with orders between $m=2$ and $m=5$ we show in panels \ref{Fig:MetaGrQNMsFull}(b)-(e) the decomposition of their intrinsic quality factor ($Q_i$) into $Q_\mathrm{rad}$ and $Q_\mathrm{res}$. More specifically, we solve the eigenvalue problem with and without material loss in order to specify the total/intrinsic ($Q_i$) and radiation ($Q_\mathrm{rad}$) quality factors, respectively. The resistive quality factor is then calculated as $Q_\mathrm{res} = (1/Q_i - 1/Q_\mathrm{rad})^{-1}$. For a more in-depth discussion on this approach and on the calculation of quality factors in general, the reader is referred to \cite{Christopoulos:2019OpEx}.
    
    As stated in the manuscript, the angle of incidence $\vartheta$ does not affect each QNM in the same way. In terms of the real part of the eigenvalue, the resonance frequency of symmetric QNMs is consistently blueshifted while the opposite is true for antisymmetric QNMs (the $x=0$ plane is used as the symmetry plane; symmetric QNMs acquire odd $m$ indeces while an even $m$ is ascribed to antisymmetric modes). As for the imaginary part, the behavior is quite more involved. 
    The component that is primarily affected by $\vartheta$ is $Q_\mathrm{rad}$, given that $Q_\mathrm{res}$ is practically constant for the QNMs due to the high field confinement on graphene. 
    Antisymmetric QNMs of order $m\geq 4$ exhibit practically no radiation ($Q_\mathrm{rad} \rightarrow \infty$) and concequently the total quality factor remains unaffected by $\vartheta$. Note that the practical absence of radiation makes these modes hard to excite (dark modes). 
    The only exception is the lowest order antisymmetric mode, $m=2$, which can become quite radiative as the incidence angle increases [see Fig.~\ref{Fig:MetaGrQNMsFull}(b)], falling even below $Q_\mathrm{res}$ (not shown). 
    
    On the contrary, symmetric QNMs have a finite $Q_\mathrm{rad}$ (they are ``bright''), with the exact value depending on the substrate height.
    The angle of incidence can lead to either an increase [$m=5$, Fig.~\ref{Fig:MetaGrQNMsFull}(e)] or decrease [$m=3$, Fig.~\ref{Fig:MetaGrQNMsFull}(c)] of $Q_\mathrm{rad}$. Thus, the total quality factor can shift either way. Note that this is not true for a metasurface with an infinite substrate (no metal reflector), see Sec.~\ref{subsec:QNMsSpectumSubstrate}.
    Finally, from the quality factor decomposition of the $m=3$ order QNM, it can be seen that resistive and radiation quality factors are almost equal ($Q_\mathrm{res} \simeq Q_\mathrm{rad} = 1\,850$). This condition is called critical coupling and leads to the perfect absorption of the incident field \cite{Tsilipakos:2021apl} when $\vartheta = 0^\mathrm{o}$. As $\vartheta$ changes, the critical-coupling condition is broken but it can be restored for any specific angle by modifying one or more geometric parameters of the resonator. 

\subsection{\label{subsec:QNMsSpectumSubstrate} Effect of the finite metal-backed substrate}

\begin{figure}[!t]
        \centering
        \includegraphics{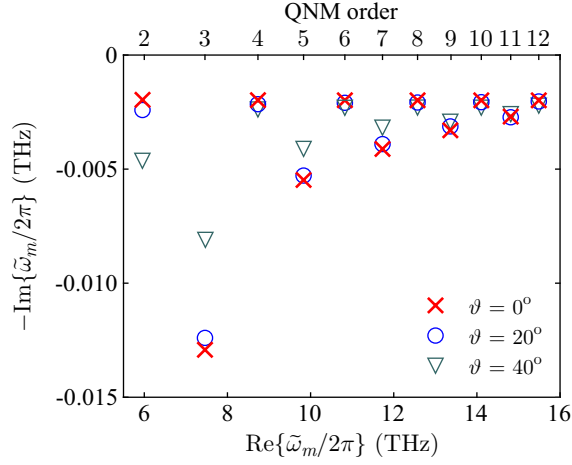}
        \caption{Complex resonance frequency of a few QNMs for a graphene-strip metasurface with infinite substrate (no metal reflector). In this example, a monotonic increase(decrease) of $Q_i$ is observed for symmetric(antisymmetric) QNMs when the angle of incidence $\vartheta$ is increased. Note that $Q_i \propto 1/|\mathrm{Im}\{\tilde\omega_m\}|$.}
        \label{Fig:MetaGrQNMsInftySubstrate}
    \end{figure}

    We have seen that in the reflective metasurface we get a rich, non-monotonic dependence of the quality factor on the angle of incidence $\vartheta$. We have attributed this behavior to the finite height of the substrate backed by the metal reflector. In order to support this claim, in Fig.~\ref{Fig:MetaGrQNMsInftySubstrate} we present the resonance frequencies of a few QNMs for a metasurface with the same graphene length and pitch, but assuming an infinite glass substrate (no metal reflector). In this case,  both symmetric and antisymmetric QNMs exhibit a monotonic $Q$-factor dependence on the incident angle. The quality factor of antisymmetric QNMs decreases with $\vartheta$ while symmetric QNMs exhibit the opposite behavior. Furthermore, symmetric QNMs of low order acquire significantly lower $Q_i$ due to their significant radiation, as also noted in Sec.~\ref{subsec:QNMsSpectumQreconstruction}. 
    As the mode order increases, the intrinsic quality factor of both symmetric and antisymmetric QNMs asymptotically approaches the limit set by $Q_\mathrm{res}$.

\subsection{\label{subsec:LinearExavulationMeta} Presentation of the $\mathbf{e}$-field envelope formulation and evaluation of the linear framework}

    In this section, we will justify the necessity of adopting a formulation  with respect to the periodic envelope ($\mathbf{e}$-field envelope formulation), in order to correctly calculate the QNMs of any resonant metasurface under oblique incidence. This stems from the implementation of the Bloch-Floquet periodic condition where the periodicity vector $\mathbf{k}_F$ needs to be specified. However, $|\mathbf{k}_F|$ depends on the unknown resonant frequency. 
    This complication can be lifted by expressing the full field $\mathbf{E}(\mathbf{r})$ using the Bloch-Floquet theorem, which states that $\mathbf{E}(\mathbf{r}) = \mathbf{e}(\mathbf{r})\exp\{-j\mathbf{k}_F\cdot\mathbf{r}\} = \mathbf{e}(\mathbf{r})\exp\{-jk_0\boldsymbol{\eta}\cdot\mathbf{r}\}$, where $k_0 = \omega/c$ is now scalar and the vector $\boldsymbol{\eta}$ carries the information of the incident wave direction and the electromagnetic properties of the semi-infinite space on the side of incidence; for instance, in a metasurface with periodicity along the $x$ axis and under illumination with an incident angle $\vartheta$,  $\boldsymbol{\eta} = n_i \sin\vartheta \mathbf{\hat x}$. This notation allows the inclusion of the term $\omega/c$ within the modified wave equation and not as a surface contribution in the respective periodic condition. More specifically, we start from the source-free Helmholtz equation for the full field in an eigenvector notation
    \begin{equation}
        \nabla\times\mu^{-1}\nabla\times\mathbf{\tilde E}_m - \tilde\omega_m^2\varepsilon\mathbf{\tilde E}_m - \tilde\omega_m\mathbf{\tilde J}_m = 0.
        \label{Eq:HelmholtzFull}
    \end{equation}
    We then introduce the Bloch-Floquet theorem and after some algebra arrive at the respective equation for the envelope $\mathbf{e}(\mathbf{r})$ \cite{Davanco:2007,Fietz:2011}
    \begin{align}
        \nabla\times\mu^{-1}\nabla\times\mathbf{\tilde e}_m &- j\frac{\tilde\omega_m}{c_0}\nabla\times\mu^{-1}(\boldsymbol{\eta}\times\mathbf{\tilde e}_m) - j\frac{\tilde\omega_m}{c_0}\boldsymbol{\eta}\times\mu^{-1} \nabla\times\mathbf{\tilde e}_m  -        \frac{\tilde\omega_m^2}{c_0^2}\boldsymbol{\eta}\times\mu^{-1}\boldsymbol{\eta}\times\mathbf{\tilde e}_m \nonumber \\ 
        &-\tilde\omega_m^2\varepsilon\mathbf{\tilde e}_m - \tilde\omega_m\mathbf{\tilde j}_m = 0.
        \label{Eq:HelmholtzEnvelope}
    \end{align}
    Equation~\eqref{Eq:HelmholtzEnvelope} is the modified wave equation for the envelope; all the required information about the direction of the incident wave and the periodicity is included within the equation through $\boldsymbol{\eta}$. 
    Since the envelope is a spatially periodic quantity and thus continuous in the boundaries of the unit cell, no Bloch-Floquet periodic boundary conditions are required; rather, a simple continuity condition is applied. Equation~\eqref{Eq:HelmholtzEnvelope} is accompanied by the auxiliary field equation for graphene
    \begin{equation}
        (\tilde\omega_m - j\gamma)\mathbf{\tilde j}_m + \dbar{\sigma}_0\mathbf{\tilde e}_m\delta_s = 0.
        \label{Eq:HelmholtzEnvelopeAux}
    \end{equation}
    
    \begin{figure}[t!]
        \centering
        \includegraphics{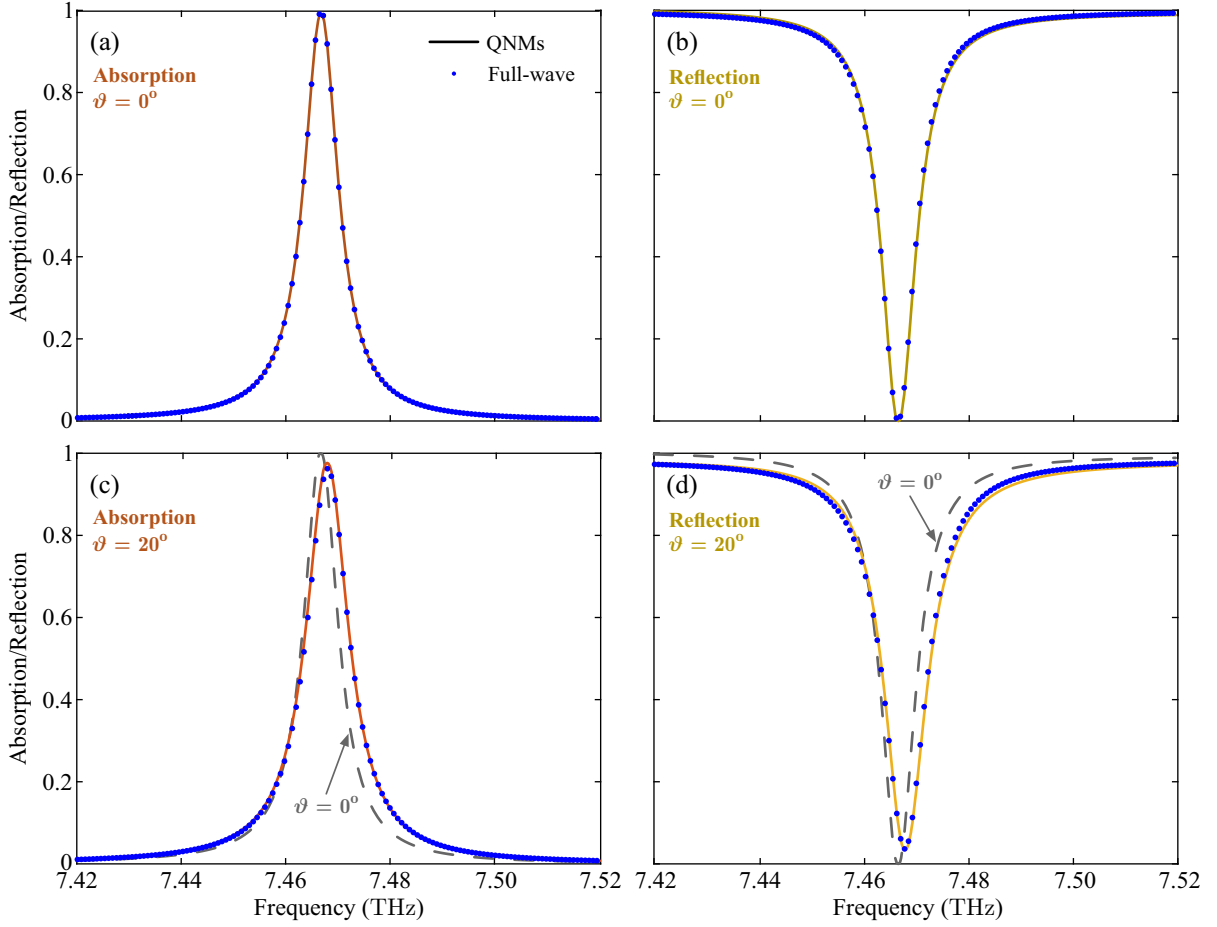}
        \caption{Absorption and reflection of the graphene strip metasurface for (a,b) $\vartheta = 0^\mathrm{o}$ and (c,d) $\vartheta = 20^\mathrm{o}$. Full-wave results (dot markers) are compared with the QNM expansion (solid lines). In (c,d) the curves of panels (a,b) are included with dashed lines for easier comparison. The agreement is excellent in all the cases considered. For $\vartheta = 20^\mathrm{o}$, the predicted blueshift of the resonance, the broadening of the Lorentzian, and braking of the critical coupling are all accurately reproduced by the framework.}
        \label{Fig:MetaGrLinearComparison}
    \end{figure}
    
    Equation~\eqref{Eq:HelmholtzEnvelope} can be solved by the generic FEM eigensolver of COMSOL  to get the correct QNMs for this example. Using these eigenmodes we can evaluate the proposed framework in the linear regime for different angles $\vartheta = 0^\mathrm{o}$ and $20^\mathrm{o}$. The results are depicted in Fig.~\ref{Fig:MetaGrLinearComparison}, revealing once again very good agreement with full-wave simulations.
    Absorption is calculated through
    \begin{equation}
        |A|^2 = \frac{1}{2P_\mathrm{inc}} \int_\ell \mathrm{Re}\{\sigma_s(\omega)\}|\mathbf{e}_\mathrm{sct,\parallel}|^2 \mathrm{d}\ell = \frac{1}{2P_\mathrm{inc}} \int_\ell \frac{\gamma}{\sigma_0} |\mathbf{j}_\mathrm{sct,\parallel}|^2 \mathrm{d}\ell,
        \label{Eq:AQNM}
    \end{equation}
    while to calculate the reflection one inevitably has to use the field outside the resonator and evaluate the expression 
    \begin{equation}
        |R|^2 = \frac{1}{2P_\mathrm{inc}} \int_\ell (\mathbf{e}_\mathrm{sct} + \mathbf{e}_r)\times(\mathbf{h}_\mathrm{sct}^* + \mathbf{h}_r^*) \mathrm{d}\ell,
        \label{Eq:RQNM}
    \end{equation}
    where $\boldsymbol{\psi}_r$ is the reflected field in the absence of the resonator and $P_\mathrm{inc} = I_0\Lambda$ is the incident wave power (in $\mathrm{W/m}$). Note that the completeness of the expansion is not strictly satisfied outside the resonant cavity, which justifies any minor discrepancy with the full-wave results. 
    As noted earlier, for $\vartheta = 20^\mathrm{o}$ the critical coupling condition will seize to be exactly satisfied and thus absorption is not perfect, compare Fig.~\ref{Fig:MetaGrLinearComparison}(c) with Fig.~\ref{Fig:MetaGrLinearComparison}(a). The predicted blueshift of the resonance frequency and the broadening of the Lorentzian due to the lower $Q$ are confirmed by the proposed framework.

    As a last remark, we note that the use of the right (which are readily available) instead of the left eigenvectors in the normalization would have introduced a significant error in the absorption and, more importantly, in the reflection spectrum due to the incorrect superposition of the background field with the field scattered by the resonator. The reader is referred to \cite{Zhang:2020} for a more in depth discussion.
    
    
    \subsection{Absorbed power as a function of incident power}
    
\begin{figure}[!t]
        \centering
        \includegraphics{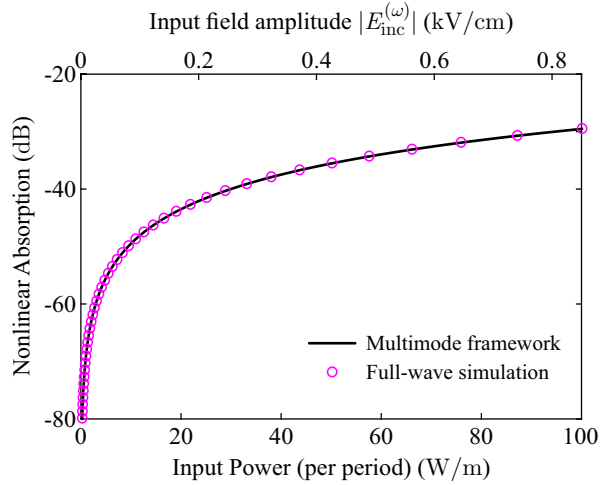}
        \caption{Nonlinear absorption of light at the third harmonic frequency by graphene. Only a small portion of the converted light is absorbed (approximately 5\%) while the rest is emitted by the metasurface (compare with  Fig.~\ref{Fig:MetaGrNLCE}).}
        \label{Fig:MetaGrNLAbsorption}
    \end{figure}

    Finally, in Fig.~\ref{Fig:MetaGrNLAbsorption} we include the nonlinear absorption of the metasurface versus input power. This plot complements Fig.~\ref{Fig:MetaGrNLCE} and reveals that only a small portion of the converted light at the third harmonic frequency  (around 5\%) is absorbed by graphene; the rest is emitted in the form of a plane wave and collected in order to calculate the conversion efficiency.
    
\bibliographystyle{ieeetr}
\bibliography{Bibliography}


\end{document}